\documentclass[12pt,tightenlines,superscriptaddress]{revtex4-2} 
\usepackage[utf8]{inputenc}

\usepackage[page,title]{appendix}
\usepackage{url}
\usepackage{placeins}
\usepackage{amsfonts}
\usepackage[dvipsnames]{xcolor}
\usepackage{color}
\usepackage{xspace}
\usepackage{tikz}
\usepackage[utf8]{inputenc}											
\usepackage[caption=false]{subfig}
\usepackage{wrapfig}

\usetikzlibrary{arrows}
\usetikzlibrary{shapes}
\usetikzlibrary{trees}
\usetikzlibrary{matrix}
\usetikzlibrary{arrows} 			
\renewcommand{\draft}[1]{{\color{blue}#1}}

\usepackage{graphicx,psfrag,bm}
\usepackage{mathrsfs,amsfonts}
\usepackage{epsfig,cancel}
\usepackage{epstopdf}
\usepackage{mciteplus}
\usepackage{latexsym}
\usepackage{amsthm}
\usepackage{amsmath}
\usepackage{amssymb}
\usepackage{hepunits}
\usepackage{hyperref}
\usepackage{bbm}
\usepackage{xfrac}
\usepackage{colordvi}
\usepackage{comment}
\usepackage{dcolumn}
\usepackage{times,wrapfig}
\usepackage{slashed}
\usepackage{booktabs}
\usepackage{relsize}
\newcommand{\be}{\begin{eqnarray}}
\newcommand{\ee}{\end{eqnarray}}

\newcommand{\benum}{\begin{enumerate}}
\newcommand{\eenum}{\end{enumerate}}
\newcommand{\bi}{\begin{itemize}}
\newcommand{\ei}{\end{itemize}}

\newcommand{\pt}{\ensuremath{p_{\text{T}}}}

\def\babar{\mbox{\slshape B\kern-0.1em{\smaller A}\kern-0.1emB\kern-0.1em{\smaller A\kern-0.2em R}}}

\newcommand{\ednote}[1]{} 
\newcommand{\people}[1]{} 
\newcommand{\morepeople}[1]{} 

\colorlet{RED}{red}
\colorlet{BLUE}{blue}
\colorlet{ORANGE}{orange}
\definecolor{darkpink}{rgb}{0.97, 0.56, 0.65}

\newcommand{\ch}[1]{{\bf\color{red} CH: #1}}

\begin{document}

\title{Current Status and Future Prospects for the Light Dark Matter eXperiment}
\date{March 15, 2022}
\revised{August 22, 2023}
\hspace*{0pt}\hfill 
{\small }

\bigskip
\author{Torsten~Åkesson}
\affiliation{Lund University, Department of Physics, Box 118, 221 00 Lund, Sweden}

\author{Nikita~Blinov}
\affiliation{University of Victoria, Victoria, BC V8P 5C2, Canada}

\author{Lukas~Brand-Baugher}
\affiliation{Rensselaer Polytechnic Institute, Troy, NY 12180, USA}

\author{Cameron~Bravo}
\affiliation{SLAC National Accelerator Laboratory, Menlo Park, CA 94025, USA}

\author{Lene~Kristian~Bryngemark}
\affiliation{Stanford University, Stanford, CA 94305, USA}

\author{Pierfrancesco~Butti}
\affiliation{SLAC National Accelerator Laboratory, Menlo Park, CA 94025, USA}




\author{Caterina~Doglioni}
\affiliation{Lund University, Department of Physics, Box 118, 221 00 Lund, Sweden}

\author{Craig~Dukes}
\affiliation{University of Virginia, Charlottesville, VA 22904, USA}

\author{Valentina~Dutta}
\affiliation{University of California at Santa Barbara, Santa Barbara, CA 93106, USA}

\author{Bertrand~Echenard}
\affiliation{California Institute of Technology, Pasadena, CA 91125, USA}

\author{Ralf~Ehrlich}
\affiliation{University of Virginia, Charlottesville, VA 22904, USA}

\author{Thomas~Eichlersmith}
\affiliation{University of Minnesota, Minneapolis, MN 55455, USA}

\author{Andrew~Furmanski}
\affiliation{University of Minnesota, Minneapolis, MN 55455, USA}

\author{Chloe~Greenstein}
\affiliation{Lewis \& Clark College, Portland, OR 97219, USA}

\author{Craig~Group}
\affiliation{University of Virginia, Charlottesville, VA 22904, USA}


\author{Niramay~Gogate}
\affiliation{Texas Tech University, Lubbock, TX 79409, USA}

\author{Vinay~Hegde}
\affiliation{Texas Tech University, Lubbock, TX 79409, USA}

\author{Christian~Herwig}
\affiliation{Fermi National Accelerator Laboratory, Batavia, IL 60510, USA}

\author{David~G.~Hitlin}
\affiliation{California Institute of Technology, Pasadena, CA 91125, USA}

\author{Duc~Hoang}
\affiliation{Massachusetts Institute of Technology, Cambridge, MA 02139, USA}

\author{Tyler~Horoho}
\affiliation{University of Virginia, Charlottesville, VA 22904, USA}

\author{Joseph~Incandela}
\affiliation{University of California at Santa Barbara, Santa Barbara, CA 93106, USA}



\author{Wesley~Ketchum}
\affiliation{Fermi National Accelerator Laboratory, Batavia, IL 60510, USA}

\author{Gordan~Krnjaic}
\affiliation{Fermi National Accelerator Laboratory, Batavia, IL 60510, USA}


\author{Amina~Li}
\affiliation{University of California at Santa Barbara, Santa Barbara, CA 93106, USA}

\author{Shirley~Li}
\affiliation{Fermi National Accelerator Laboratory, Batavia, IL 60510, USA}

\author{Dexu~Lin}
\affiliation{California Institute of Technology, Pasadena, CA 91125, USA}


\author{Jeremiah~Mans}
\affiliation{University of Minnesota, Minneapolis, MN 55455, USA}

\author{Cristina~Mantilla~Suarez}
\affiliation{Fermi National Accelerator Laboratory, Batavia, IL 60510, USA}

\author{Phillip~Masterson}
\affiliation{University of California at Santa Barbara, Santa Barbara, CA 93106, USA}



\author{Martin~Meier}
\affiliation{University of Minnesota, Minneapolis, MN 55455, USA}

\author{Sophie~Middleton}
\affiliation{California Institute of Technology, Pasadena, CA 91125, USA}

\author{Omar~Moreno}
\affiliation{SLAC National Accelerator Laboratory, Menlo Park, CA 94025, USA}

\author{Geoffrey~Mullier}
\affiliation{Lund University, Department of Physics, Box 118, 221 00 Lund, Sweden}


\author{Timothy~Nelson}
\affiliation{SLAC National Accelerator Laboratory, Menlo Park, CA 94025, USA}

\author{James Oyang}
\affiliation{California Institute of Technology, Pasadena, CA 91125, USA}


\author{Jessica~Pascadlo}
\affiliation{University of Virginia, Charlottesville, VA 22904, USA}

\author{Ruth~Pöttgen}
\affiliation{Lund University, Department of Physics, Box 118, 221 00 Lund, Sweden}

\author{Stefan~Prestel}
\affiliation{Lund University, Department of Physics, Box 118, 221 00 Lund, Sweden}

\author{Luis~Sarmiento~Pico}
\affiliation{Lund University, Department of Physics, Box 118, 221 00 Lund, Sweden}

\author{Philip~Schuster}
\affiliation{SLAC National Accelerator Laboratory, Menlo Park, CA 94025, USA}

\author{Matthew~Solt}
\affiliation{University of Virginia, Charlottesville, VA 22904, USA}

\author{Lauren~Tompkins}
\affiliation{Stanford University, Stanford, CA 94305, USA}

\author{Natalia~Toro}
\affiliation{SLAC National Accelerator Laboratory, Menlo Park, CA 94025, USA}

\author{Nhan~Tran}
\affiliation{Fermi National Accelerator Laboratory, Batavia, IL 60510, USA}

\author{Andrew~Whitbeck}
\affiliation{Texas Tech University, Lubbock, TX 79409, USA}

\author{Kevin~Zhou}
\affiliation{SLAC National Accelerator Laboratory, Menlo Park, CA 94025, USA}

\author{Laura~Zichi}
\affiliation{University of Michigan, Ann Arbor, MI 48109, USA}


\begin{abstract}
The constituents of dark matter are still unknown, and the viable possibilities span a vast range of masses.  The physics community has established searching for sub-GeV dark matter as a high priority and identified accelerator-based experiments as an essential facet of this search strategy \cite{BRNReport,Battaglieri:2017aum}. A key goal of the accelerator-based dark matter program is testing the broad idea of thermally produced sub-GeV dark matter through experiments designed to directly produce dark matter particles. 
The most sensitive way to search for the production of light dark matter is to use a primary electron beam to produce it in fixed-target collisions. The Light Dark Matter eXperiment (LDMX) is an electron-beam fixed-target missing-momentum experiment that realizes this approach and provides unique sensitivity to light dark matter in the sub-GeV range.  This contribution provides an overview of the theoretical motivation, the main experimental challenges, how LDMX addresses these challenges, and projected sensitivities.  We further describe the capabilities of LDMX to explore other interesting new and standard physics, such as visibly-decaying axion and vector mediators 
or rare meson decays, and to provide timely electronuclear scattering measurements that will inform the modeling of neutrino-nucleus scattering for DUNE. 

\end{abstract}


\maketitle
\newpage
\tableofcontents
\newpage

\maketitle

\clearpage


\section{Introduction}
A predictive explanation for the origin of dark matter (DM), which is also compelling from an experimental perspective, is that of a thermal relic from the early Universe. Thermal-relic DM can explain present-day observations if the mass of the DM is between roughly 1 MeV and 100 TeV. The interactions necessary for DM to thermalize with SM particles also imply that DM production at current accelerator facilities is possible up to roughly 1 TeV masses. The sub-GeV region of this parameter space is well-motivated and a critical piece for a comprehensive test of the thermal-relic hypothesis. Details of relevant dark matter models and recent reviews can be found in~\cite{Bjorken:2009mm,Izaguirre:2015yja,Izaguirre:2014bca,Battaglieri:2017aum,Alexander:2016aln,BRNReport}.

The Light Dark Matter eXperiment (LDMX)~\cite{Akesson:2018vlm} is a fixed-target missing momentum experiment.  LDMX will leverage ongoing LCLS-II upgrades~\cite{Raubenheimer:2018mwt} to attain high enough integrated luminosities to test the thermal targets for a variety of models across a range of DM masses. The Linac to End-Station A (LESA) will deliver beam to LDMX beginning in 2025, allowing a sample of $4\times10^{14}$ electrons on target (EoT) to be collected in the initial phase of running, lasting roughly one year. A second phase is planned to begin in 2027, when a dedicated injector laser will allow LESA to deliver as many as $10^{16}$~EoT in the following few years. Beam electrons are focused upon a thin (0.1$X_{0}$) tungsten target in an attempt to produce DM. The downstream apparatus consists of a charged particle tracker and hermetic calorimeters to verify the potential DM signature: a single, low-transverse momentum (\pt) recoil electron accompanied by the complete absence of other particle activity. The requirements for this veto system are set by several key background processes including electronuclear and photonuclear interactions that transform visible energy into more difficult to detect secondary particles such as neutrons and kaons. In the event of an observation, the LDMX detector design enables the potential DM signal to be characterized through the measured missing-momentum spectrum, uncovering the relevant mass scale to inform future studies.

Though motivated by the search for light thermal relics, the design of the LDMX detector is well-suited to enable a broader physics program, shedding light on further pressing questions. The first is a broader class of DM models that predict visible and displaced signatures not covered by the missing-momentum search, including axion-like particles (ALPs), strongly-interacting dark sectors (SIMPs), milli-charged particles, and more~\cite{Berlin:2018bsc}. Another avenue aims to complement the next-generation of long-baseline experiments such as the Deep Underground Neutrino Experiment (DUNE)~\cite{DUNE:2020lwj,DUNE:2020ypp} and Hyper-Kamiokande~\cite{Hyper-Kamiokande:2018ofw,Hyper-Kamiokande:2022smq} that will use neutrino beams to better understand the nature of these particles. However, such experiments rely on models of neutrino-nucleus interactions that are difficult to constrain without precise knowledge of the incoming neutrino energy. Full-event reconstruction of electro-nuclear interactions with LDMX presents an opportunity to improve such models, particularly in the previously unexplored far-forward region~\cite{Ankowski:2019mfd}. As a last example, LDMX offers the potential to make precise measurements of rare meson decays, predicted by the Standard Model (SM) but never before measured~\cite{Schuster:2021mlr}. A number of avenues exist to upgrade LDMX beyond the design envisioned for Phase 2, extending the experiment's sensitivity to these signatures and more. 

The remainder of this manuscript is organized as follows. Section~\ref{sec:det} briefly reviews the experimental concept before reviewing recent work towards prototype demonstrations of key subdetector components. Section~\ref{sec:dm} describes the baseline search strategy for light thermal relics, including analysis of multi-electron events and methods of background estimation using data-driven techniques. A strategy to analyze early data for DM production in the ECal, as well as projected performance with an 8~GeV $e^-$ beam. Other searches for light New Physics are described in Section~\ref{sec:otherNP}, including visible and semi-visible final states. Section~\ref{sec:en} describes the potential to make precision measurements of electro-nuclear physics, enriching the potential of future neutrino beam experiments. The penultimate Section~\ref{sec:future} proposes potential avenues to further enrich the physics program of LDMX, including new beam and detectors configurations, before concluding remarks are offered in Section~\ref{sec:conc}.

\section{Detector concept and status}
\label{sec:det}
\subsection{Review of the experimental apparatus}

\begin{figure}
    \centering
    \includegraphics[width=0.95\textwidth]{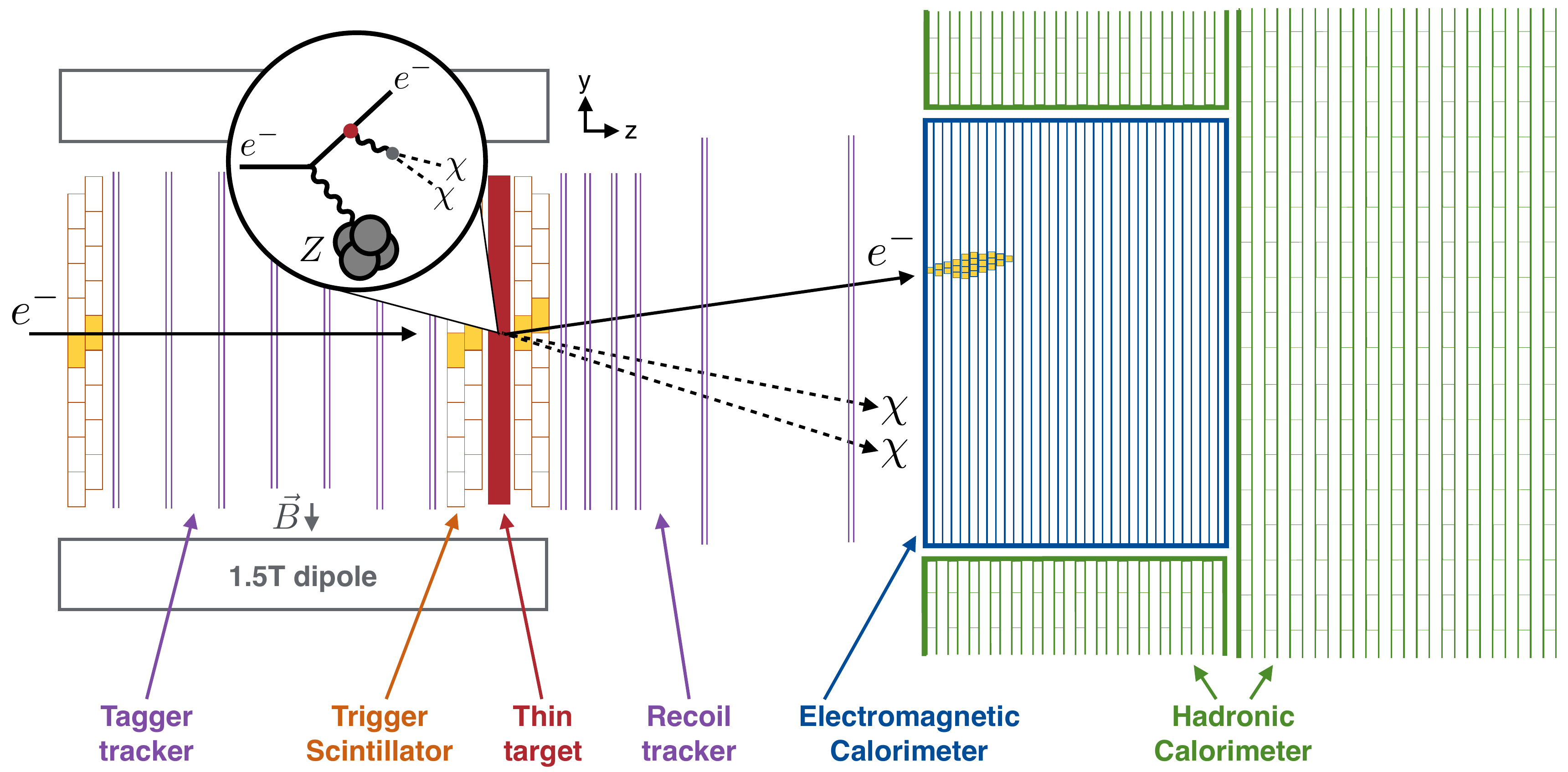}
    \caption{A diagram of the LDMX detector apparatus, illustrating production of DM in the target from a scattering beam electron, and the corresponding response of the various sub-systems to the missing-momentum signature.  The drawing is not to scale.}
    \label{fig:detector}
\end{figure}

\begin{figure}
    \centering
    \includegraphics[width=0.95\textwidth]{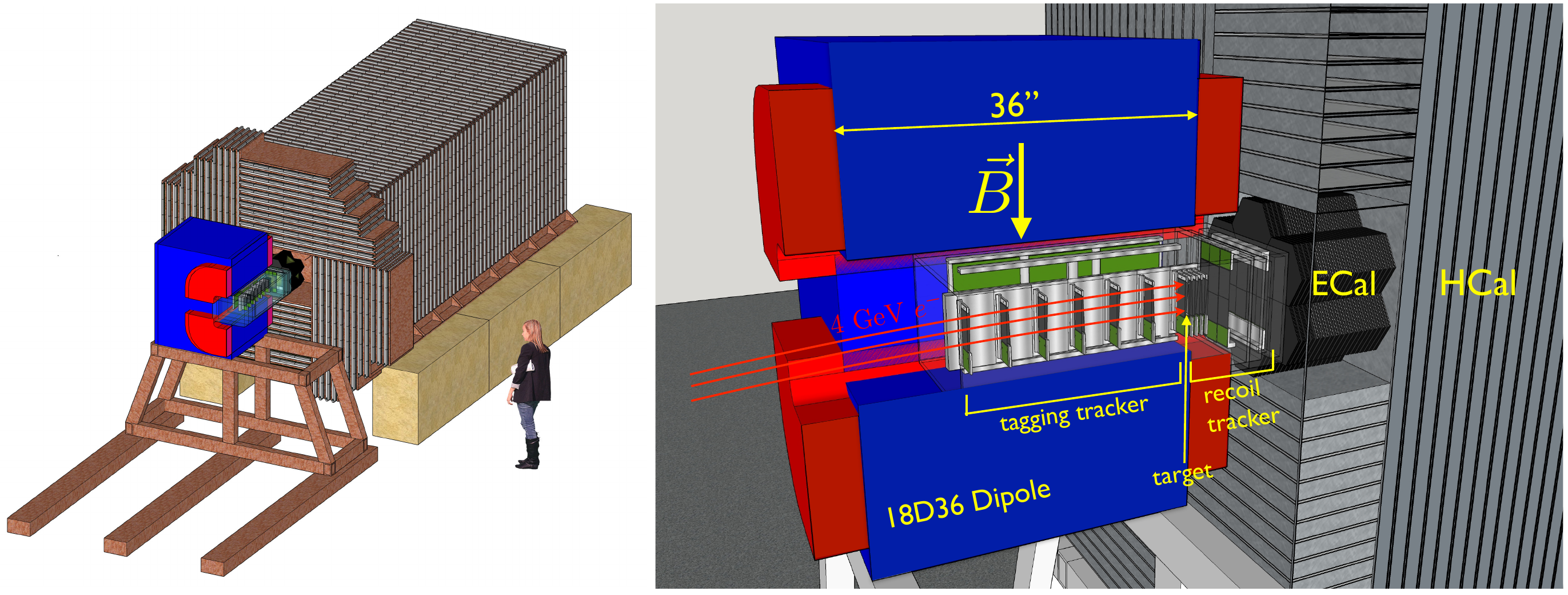}
    \caption{At left, the complete LDMX detector apparatus is shown including mechanical supports, with a human for scale.  At right, the detector has been partially cutaway to show the various sub-detectors.}
    \label{fig:detector2}
\end{figure}

The LDMX detector apparatus is designed to collect a large dataset of single-electron scatters off of a thin target, providing a complete picture of the resulting final state reconstructed from downstream particle interactions. The detector consists of a tracking system with a strong magnetic field to precisely measure electron momenta before and after passing through the target, deep electromagnetic and hadronic calorimeters to capture complete particle showers, and a trigger system to collect events of interest. The tagging tracker system of stereo silicon strip modules is situated within a 1.5\,T dipole magnet upstream of the target, nominally 0.1 $X_0$ of Tungsten. Downstream of the target, a similarly-constructed recoil tracker immersed in the fringe magnetic field reconstructs charged particles with $p\geq 50$\,MeV. Fiducial beam electrons are identified in real time by the trigger scintillator (TS), consisting of plastic or LYSO bars upstream of the tagger tracker and on either side of the target. A Si-W electromagnetic calorimeter (ECal) 40 $X_0$ in depth allows precise reconstruction of charged particle showers and provides a fast total energy measurement for trigger events in combination with the TS. A large hadronic calorimeter (HCal) consisting of scintillating bars and steel absorbers contains highly-penetrating neutral hadrons ($\sim17\lambda_I$) as well as scattering products at very wide angles. Figure~\ref{fig:detector} illustrates the LDMX detector concept and DM signature, while Figure~\ref{fig:detector2} provides a to-scale mechanical mock-up of the full apparatus.

A complete description of the LDMX detector concept was laid out in Ref.~\cite{Akesson:2018vlm}. Since that time, significant work had gone into improving the design of each sub-detector, constructing prototype demonstrations of various hardware components, and continuing to develop sophisticated simulation tools to better map out the anticipated performance of the completed detector. 

\subsection{Prototype development and testbeam demonstration}

%

Prototypes of the HCal and TS components have been constructed to validate the design of the devices in realistic conditions, and measure their response as a function of the beam energy and particle type. An in-situ calibration of the LDMX HCal time and energy resolution will be challenging, due to the configuration of the experiment. Input from the Monte Carlo (MC) simulation is likely to be required to properly estimate the device response. The modeling of hadronic interactions in the few-GeV region suffers from non-negligible uncertainties, and additional measurements in that region will be invaluable to tune the MC response and ensure an accurate detector calibration. Realistic $e^-$ beam data is critical to map the response of the TS, determine hit efficiencies, and compare scintillator technologies. The prototype detectors were developed for data-taking with a few-GeV beam at the CERN T9 line, and are essentially identical to the corresponding LDMX sub-detectors, up to size differences.

\begin{figure}[tbh]
    \centering
    \includegraphics[width=0.2\textwidth]{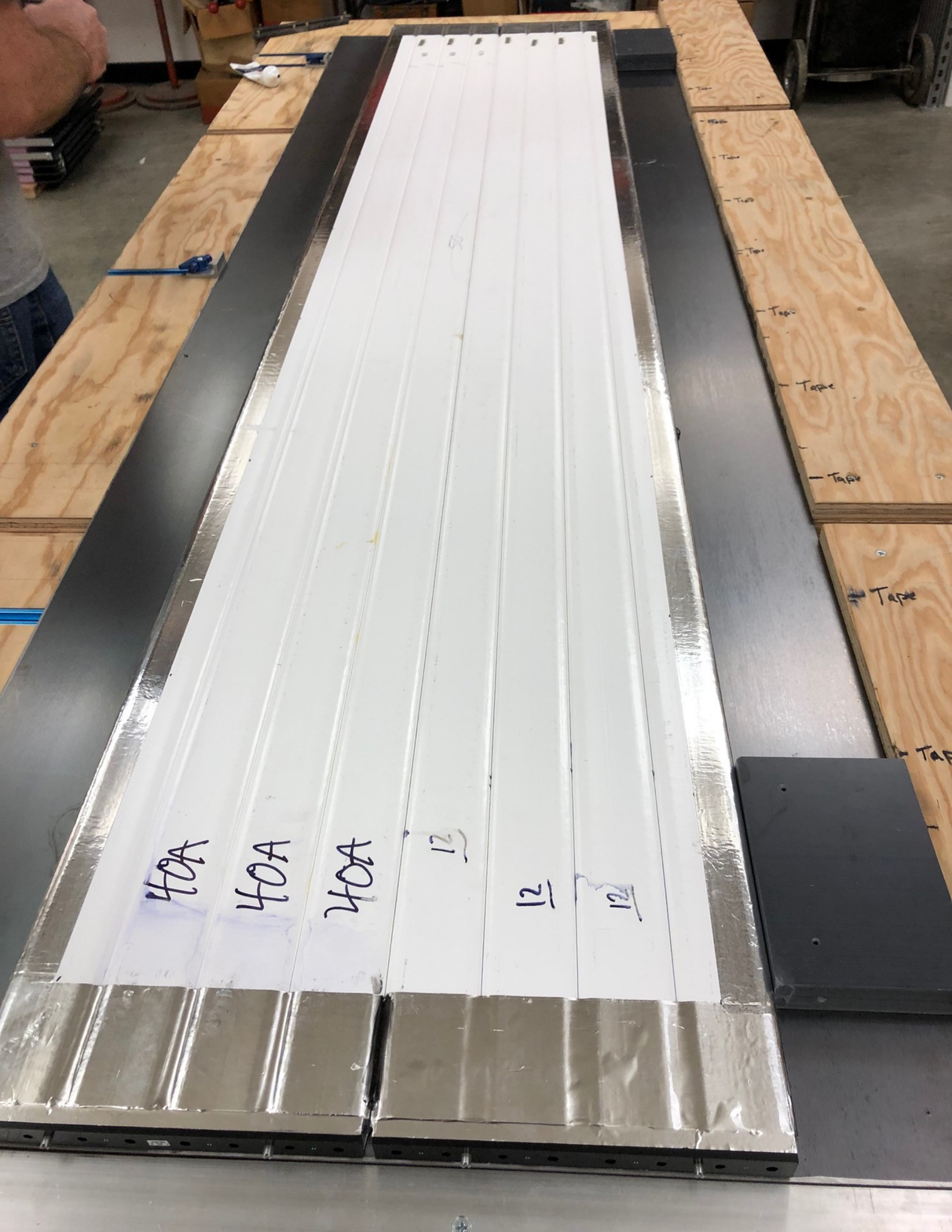}
    \includegraphics[width=0.4\textwidth]{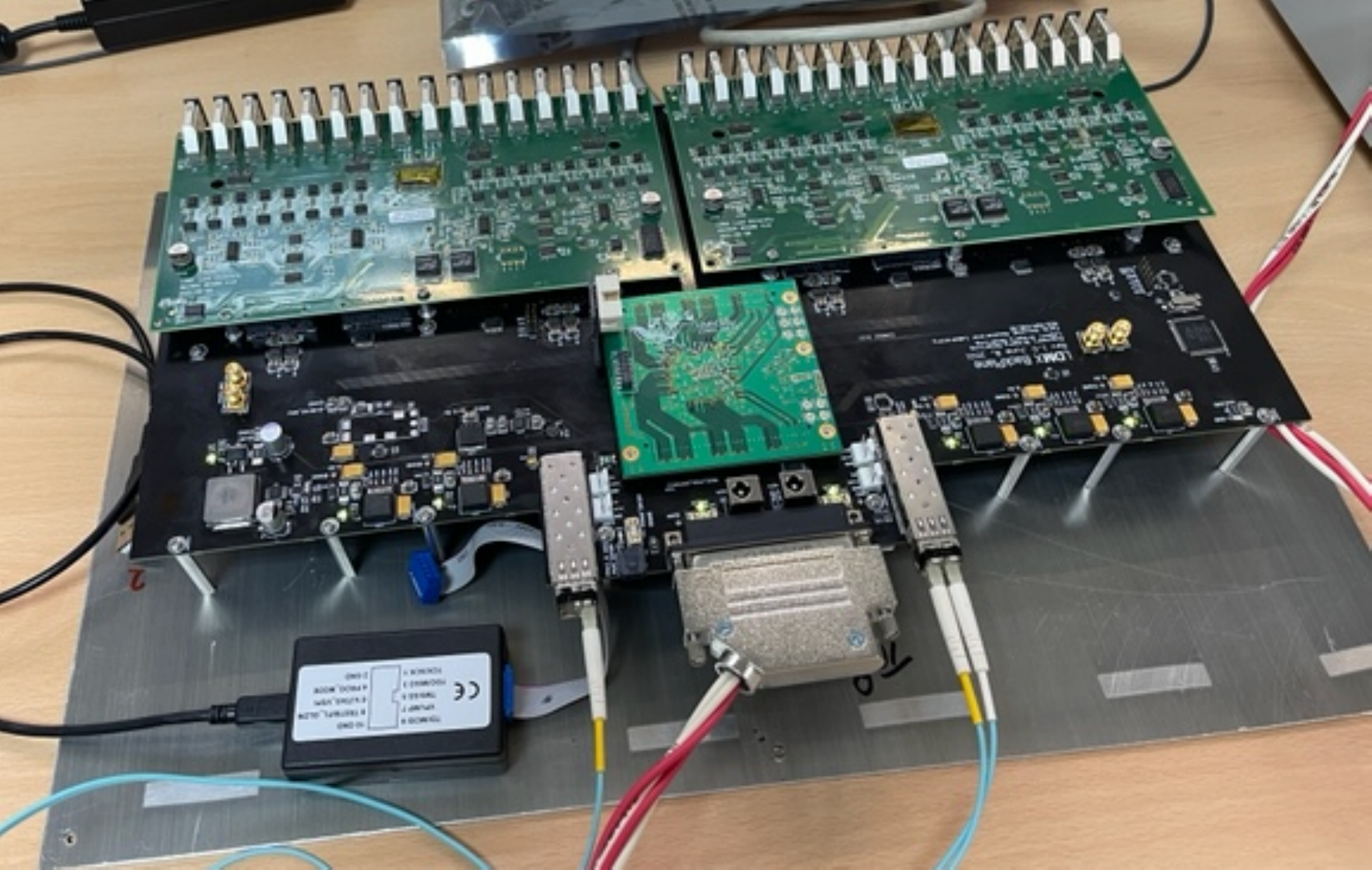}
    \includegraphics[width=0.28\textwidth]{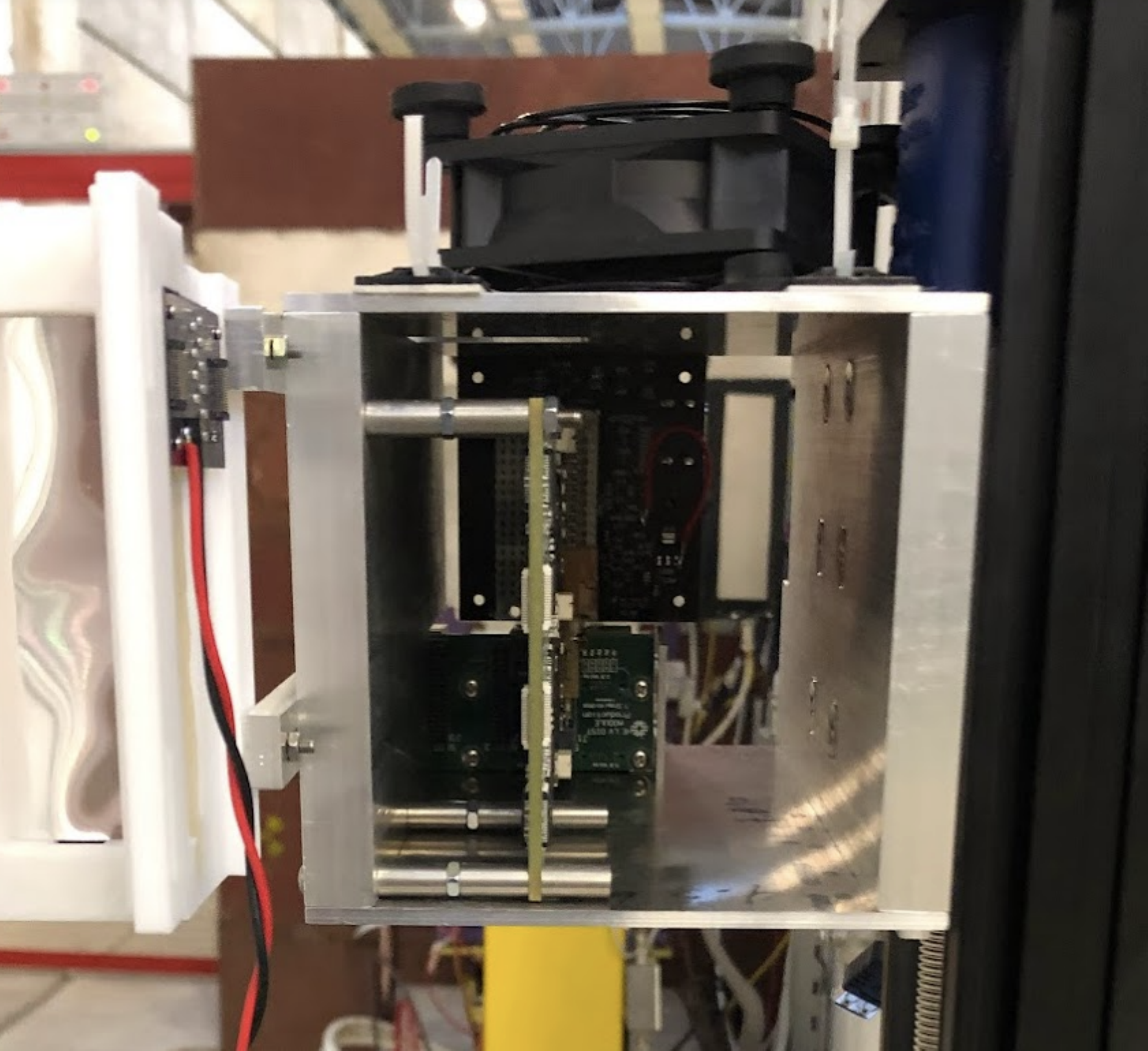}
    \caption{
    Left: A scintillator layer of the HCal prototype composed of two sets of four bars each. 
    Center: Prototype of LDMX HCal custom readout electronics which digitizes 256 SiPM signals using the HGCROCv2. Digitized signals are aggregated on an FPGA mezzanine and transmitted optically to a central ATCA-based DAQ system. 
    Right: the prototype TS module.  Scintillators are read out by 12 SiPMs, whose signals are digitized by a QIE11 hosted on a readout board, shown vertically in the aluminum housing.  Data is transmitted optically over a pair of fibers to the trigger and DAQ system.}
    \label{fig:hcalproto}
\end{figure}

\paragraph{Sampling calorimeter}
The prototype HCal consists of 19 layers of 25\,mm iron absorber and 20\,mm scintillator, corresponding to about three nuclear interaction lengths. The absorber has transverse dimensions of 70\,cm x 70\,cm, sufficient to contain typical hadronic showers. The scintillator bars are 2\,m long and 5\,cm wide and are grouped into quad-counters of 4 individually-readout bars. The first 9 layers contain 2 quad-counters, while the remaining layers have 3 quad-counters to better capture the hadronic shower. A scintillator layer is shown in Figure~\ref{fig:hcalproto}. The bars have a through-hole into which a wavelength-shifting fiber is inserted, which is read out at each end with silicon photo-multipliers (SiPMs). The layers are arranged in an $x$-$y$ configuration, and can be moved to measure the response as a function of the distance to the readout. These components rest on a movable mounting structure adjustable to take measurements at different angles with respect to the beam direction. 

\paragraph{Readout electronics}
A prototype custom readout system for the LDMX HCal was designed and commissioned in preparation for the CERN testbeam. The system is designed to take the SiPM signals described above, transmitted over HDMI, and digitize and transmit those signals to the central DAQ and trigger systems. The readout board, as shown in Fig.~\ref{fig:hcalproto}, consists of 4 ADC mezzanine cards each processing 64 SiPM channels using v2 of the HGCROC developed by the CMS experiment for readout of their future endcap calorimeter upgrade~\cite{hgcalTDR}. Those ADC mezzanine cards are powered and supported by a backplane card which also supports an FPGA mezzanine card that aggregates digitized data. The FPGA mezzanine card will be shared between the ECal and HCal systems.

For the LDMX CERN testbeam, each of the components of the electronics system were tested individually and integrated successfully. Preliminary channel pedestal calibrations were performed. The entire data readout chain from the HCal readout electronics to the central DAQ system was also demonstrated.  

\paragraph{Trigger Scintillator Prototype}
The trigger scintillator prototype consists of an array of 6 EJ-200 plastic scintillator bars that are 2x3x30 mm and LYSO bars that are 0.6x3x30 mm.  The plastic bars are arranged in a 1x6 array while the LYSO bars are arranged in a 2x7 array with each row offset by 1.5 mm. Scintillators are read out by 2x2\,mm SiPMs that were placed at the ends of bars.  Signals from SiPMs are digitized by Charge Integrator and Encoder (QIE) ASICs hosted on a CMS barrel hadronic calorimeter readout board. Data from 12 QIEs is transmitted over a pair of 5 Gbps optical cables to a CAPTAN+X board, which in turn sends data packets over ethernet to a data acquisition computer. A custom backplane was used to distribute power, clock, and slow control signals to the readout boards.

\subsection{Software and computing tools}

LDMX utilizes a \texttt{C++} event processing and simulation framework (\texttt{ldmx-sw}) built on top of tools used commonly throughout the HEP community. The data processing pipeline is steered with an embedded python interpreter, allowing for simple configuration and dynamic loading of simulation, digitization, reconstruction, and general analysis workflows. This robust infrastructure enables detailed studies of detector prototypes, development of reconstruction techniques, the design of complex physics analysis strategies, and more.

Studies of the response of the LDMX detector, particularly to rare SM processes and potential signals of new physics, are enabled by the Geant4 toolkit, including customizations to the Bertini Cascade model, improved matrix elements for muon conversions, and incorporation of dark-bremsstrahlung signal generation\cite{Akesson:2018vlm}. A complete and modular description of LDMX in GDML allows efficient studies of various detector configurations as well as testbeam-specific prototypes. Lastly, the Lightweight Distributed Computing System (LDCS)\cite{Bryngemark:2021} allows for the efficient and large-scale generation of simulated MC event samples, critical to understand the rare processes necessary for the analysis of $10^{16}$~EoT.

\section{Searches for light thermal relics}
\label{sec:dm}
\label{sec:dm:intro}
This section describes the search for light thermal DM with the missing momentum approach. We first review the benchmark signal models and production mechanisms, focusing on reactions where the DM is produced in the target via decay of a vector mediator. We then review the baseline selection criteria, designed to reject all sources of background while maintaining high signal efficiency. While these criteria were necessarily built from simulation, we also describe the first steps towards a data-driven validation of the veto-efficiencies and certain background rates. Next, we present the potential for LDMX to characterize the DM mass scale in the event of an observation. Lastly, we present a pair of analyses targeting alternate run scenarios: the prospect of producing DM within the material of the ECal, targeting the $10^{13}$~EoT expected in the earliest stages of the experiment; and, the extended sensitivity of LDMX when taking 8~GeV electron beam data and analyzing multiple electrons per time sample to accumulate $10^{16}$~EoT.

\subsection{Review of models and production mechanisms}

\begin{figure}
    \centering
    \includegraphics[width=0.95\textwidth]{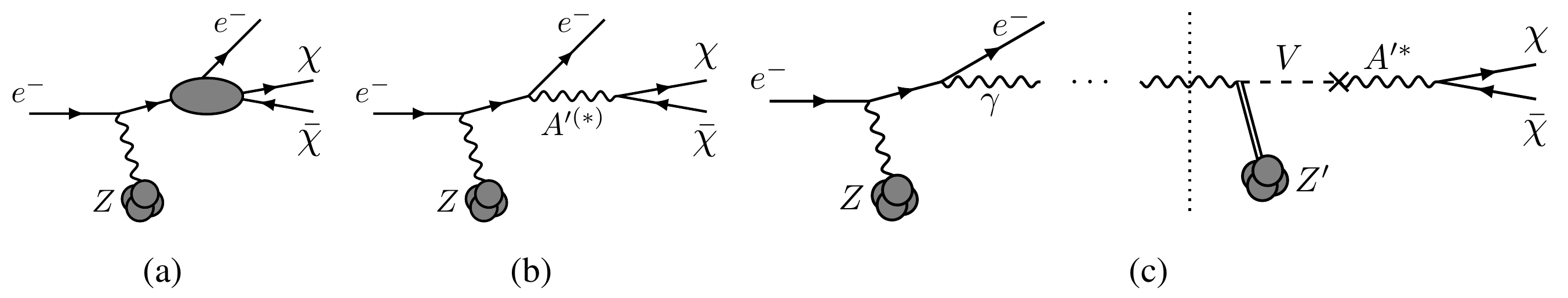}
    \caption{Schematic depiction of the DM signal at LDMX from (a) direct dark matter particle-antiparticle production, (b) $A'$ bremsstrahlung, and (c) invisible vector meson decay. For $A'$ bremsstrahlung, DM is produced through an on- or off-shell $A'$ in the target (or the calorimeter). In the meson signal process, a hard photon is produced in the target, and converts to a vector meson $V$ through an exclusive photo-production process in the calorimeter before decaying invisibly.}
    \label{fig:proddiagram}
\end{figure}

The missing momentum DM search at LDMX is sensitive to \textit{any} process in which a beam electron transfers most of its energy to invisible particles and receives an appreciable transverse kick from the production of these invisible particles. DM signals resulting in this experimental signature can be produced via:
\begin{itemize}
    \item \textit{dark bremsstrahlung}, where an electron scatters off of a nucleus and produces a pair of DM particles either directly through an effective interaction as in Fig.~\ref{fig:proddiagram}(a), or through the production and decay of a mediator particle  Fig.~\ref{fig:proddiagram}(b); or
    \item \textit{photo-production of vector mesons}, from a hard bremsstrahlung photon that scatters off of a nucleus, and later decays invisibly to dark matter particles via mixing with a mediator particle Fig.~\ref{fig:proddiagram}(c).
\end{itemize}
These production modes can be powerful probes of the DM's coupling to electrons and to quarks, with complementary strengths that depend on the details of the model considered~\cite{Schuster:2021mlr}.
An additional production channel exists, where positrons produced in electromagnetic showers via $\gamma^*\to e^+ e^-$ then annihilate with atomic electrons to produce an $A’$~\cite{Marsicano:2018glj}.
This contribution enhances experimental sensitivities in a narrow region around $\sqrt{2m_e E_\text{beam}}$ (60 MeV for the 4~GeV LDMX beam) and has not been included in the projections shown in this document.

For our benchmark scenario, we consider direct annihilation models where the population of DM $\chi$ is reduced in the early universe through annihilation $\chi\chi \rightarrow A' \rightarrow ff$ to SM fermions $f$ via a vector mediator $A'$. The specific nature of the DM $\chi$ (e.g. scalar, fermion, single- or multi-component) may vary depending on the model considered and can impact relic targets, but generally does not alter the relevant phenomenology for the missing-momentum search. 
We focus on DM produced by dark bremsstrahlung from the decay of the new $U(1)$ gauge boson $A'$, typically called the dark photon.
The gauge coupling $g_D$ parameterizes the $A'$ coupling to DM with strength $\alpha_D=g_D^2/4\pi$, while kinetic mixing with the SM photon gives rise to a weak electromagnetic interaction proportional to the mixing parameter $\epsilon$.
In the case where $m_{A'} > 2m_\chi$, the mediator is produced on-shell and decays primarily to DM.

\subsection{The 4~GeV \texorpdfstring{$e^-$}{e-} beam analysis}
\label{sec:fourGeV}

The baseline analysis for LDMX is designed for a data-taking period corresponding to $4\times10^{14}$ electrons on target (EoT), at a 4~GeV beam energy. The event selection is designed to retain high signal efficiency while aiming to be background-free, and allows LDMX to probe several key thermal targets in the 1-100 MeV mass range during this baseline run. A detailed discussion of the analysis can be found in Ref.~\cite{Akesson:2019iul}, which includes results for the background rejection performance based on studies carried out with high-statistics samples of simulated events.

The DM production signal is characterized by a low-energy recoil electron, with DM carrying away the majority of the beam energy. We select events with recoil electron energy $<1.2$~GeV. The leading source of low-energy electrons in the experiment, which can constitute a background to a DM signal, comes from events in which the incoming electron undergoes a hard bremsstrahlung in the target, producing a multi-GeV photon. Typically, such photons initiate electromagnetic showers that deposit a large amount of energy in the ECal. However, in rare cases, processes such as photonuclear (PN) interactions or $\gamma^*\to\mu^+\mu^-$ conversions can occur in the target or ECal, resulting in low energy deposition in the ECal. Depending on the interaction and the secondary particles that are produced, these backgrounds may result in a distinctive spatial profile of energy deposition in the ECal, while also producing observable hits in the HCal and/or recoil tracker. Analogous electroproduction reactions mediated by virtual photons may also occur, e.g. electronuclear scattering and muon trident production. The rates of these reactions are suppressed with respect to the corresponding real photon-induced process. They may also lead to the presence of multiple charged tracks in the recoil tracker, providing additional veto handles.

\begin{figure}
    \centering
    \includegraphics[width=0.7\textwidth]{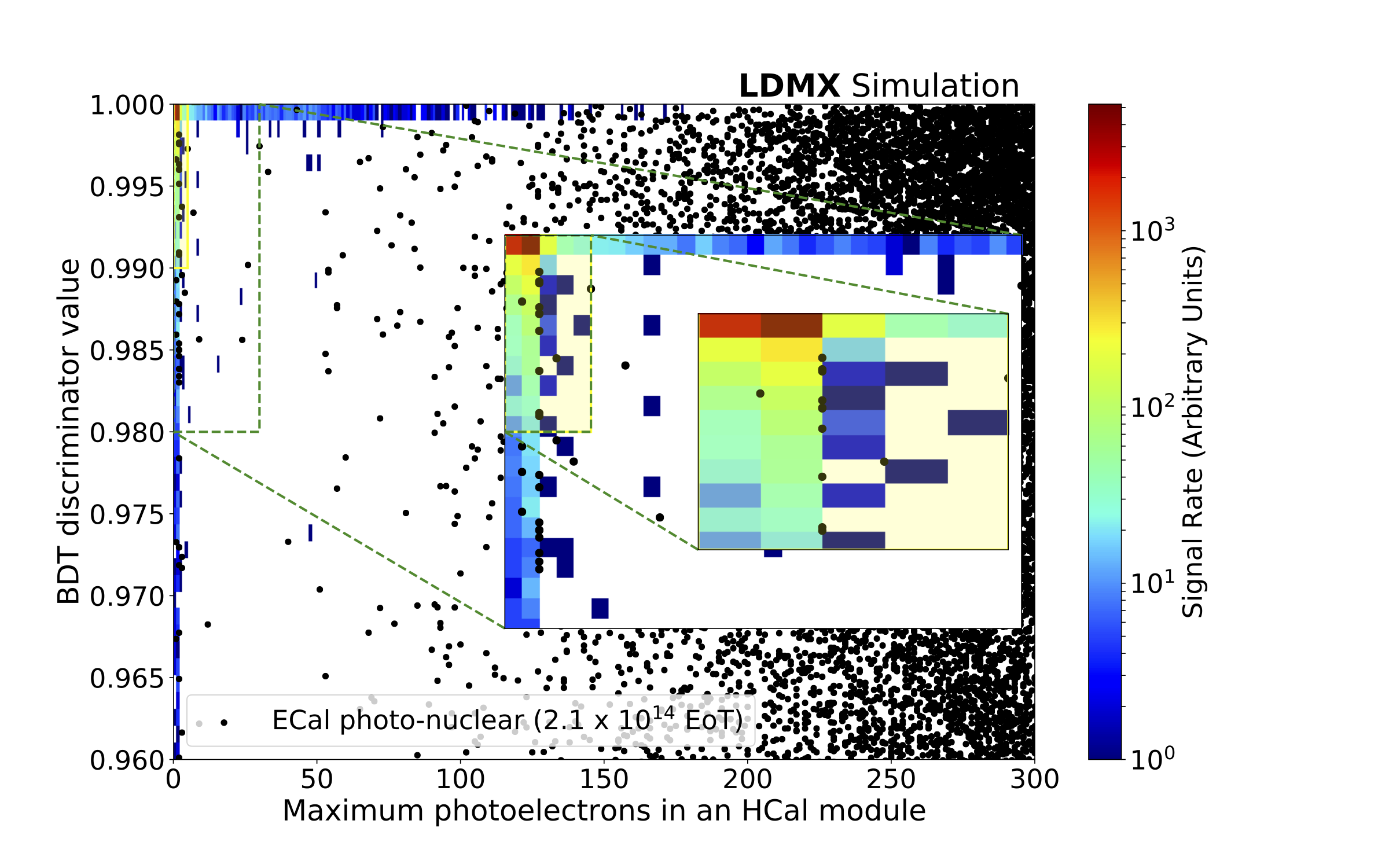}
    \caption{Distribution of ECal BDT discriminator value versus the maximum number of PEs in an HCal scintillating bar for a $2.1\times10^{14}$~EoT equivalent ECal-photonuclear sample (black points) and a signal sample with an $A'$ mass of 100 MeV (heat map). The shaded yellow box indicates the signal region. ECal photonuclear background events surviving the ECal BDT and HCal vetoes can be rejected by identifying track features in the ECal, as described in the text.
    Figure is reproduced from Ref.~\cite{Akesson:2019iul}.}
    \label{fig:bdt_maxpe}
\end{figure}

The baseline analysis comprises the following selection criteria:
\begin{description}
    \item[Missing energy trigger] Signal events would be characterized by a low-energy recoil electron and no other visible final-state particles, resulting in low energy deposition in the ECal. In contrast, background events would typically have most of the beam energy converted into electromagnetic showers in the ECal. The primary trigger strategy for LDMX therefore relies on imposing an upper threshold on the energy reconstructed in the ECal. The trigger scintillator system provides information about the number of total incoming beam electrons in a time sample, which in turn determines the missing energy threshold for the trigger. In single-electron events, the threshold is 1.5~GeV, requiring the missing energy to be at least 2.5~GeV for events to pass the trigger. A corresponding offline selection, requiring the energy reconstructed with the full Ecal to be less than 1.5~GeV, is also imposed.
    \item[Track selection] In order to select events where the incoming electron undergoes significant energy loss in the target, we require a single track in the recoil tracker with $p < 1.2$~GeV. The requirement of a single track helps to suppress backgrounds with a hard bremsstrahlung photon that interacts in the target (e.g. by converting to electrons or muons, or undergoing photonuclear scattering) and may produce additional tracks, as well as electronuclear reactions in the target that also typically produce multiple tracks. The dominant background that survives the track selection and missing energy requirements is comprised of events in which the electron undergoes a hard bremsstrahlung in the target, and the emitted photon subsequently undergoes a photonuclear reaction or conversion to muons either in the target or the ECal.
    \item[ECal boosted decision tree] A number of features based on information about energy deposition in the ECal are used to construct a multi-variate boosted decision tree (BDT) that provides powerful rejection for events with photonuclear interactions in the ECal, as well as backgrounds with target photonuclear interactions or muon conversions that survive the previous selection criteria. The variables used in the BDT include global variables that are constructed from sums or averages over the whole ECal, for example, the total energy reconstructed in the ECal. Information about the transverse and longitudinal shower profile in the ECal is also exploited. Photonuclear events are typically characterized by energy deposition deeper in the ECal compared to signal events, and may also have a broader transverse shower profile than signal events as a consequence of energy depositions from the photonuclear products. Finally, the high granularity of the ECal, which makes it possible to separate multiple showers, is also exploited by defining regions that are expected to contain most of the shower energy from the recoil electron as well as a bremsstrahlung photon under the background hypothesis. Features related to energy deposition in these ``containment regions'' are included in the BDT, and provide discriminating power since signal events are expected to have significantly less activity in the photon containment regions than background. The BDT discriminant is explicitly designed to be absent of any correlation with the recoil electron \pt.
    \item[HCal veto] Hits in the HCal provide a complementary handle to the ECal BDT for the rejection of photonuclear background as well as muon conversions. Hadronic showers originating from the products of photonuclear reactions, or muons penetrating into the HCal will typically leave a large number of hits. The HCal is also the main veto handle for photonuclear events in which the majority of the photon's energy is carried away by a single, energetic, forward-going neutral hadron such as a neutron or $K^0_\textrm{L}$. The HCal design has therefore been optimized to efficiently veto these energetic neutral hadrons. For the purposes of this baseline analysis, a veto-able hit is defined as the production of at least 5 photoelectrons (PEs) in a single HCal scintillator bar in-time with the beam electron. The HCal veto requires that there be no such hit in the entire HCal. Figure~\ref{fig:bdt_maxpe} displays the distribution of the ECal BDT score vs the maximum PEs in any HCal scintillating bar for a simulated sample of ECal photonuclear events as well as for a representative signal sample. The signal region is defined by a BDT score $>0.99$ and $<5$ for the maximum number of PEs in a bar.
    \item[ECal track features] The remaining background surviving the preceding selection criteria arises mainly from photonuclear reactions that produce a $K^\pm$ decaying in flight in the ECal in which most of the kaon energy is transferred to a neutrino, and a short track may be the only visible distinguishing feature in the ECal. The high granularity of the ECal makes it possible to identify tracks from minimum ionizing particles (MIPs) traversing the ECal. A dedicated algorithm identifies tracks or isolated hits from charged particles originating from photon interactions in the ECal, and has been found to be efficient in rejecting the remaining photonuclear background with charged kaons that survives the ECal BDT and HCal vetoes.
\end{description}

\begin{figure}
    \centering
    \includegraphics[width=0.8\textwidth]{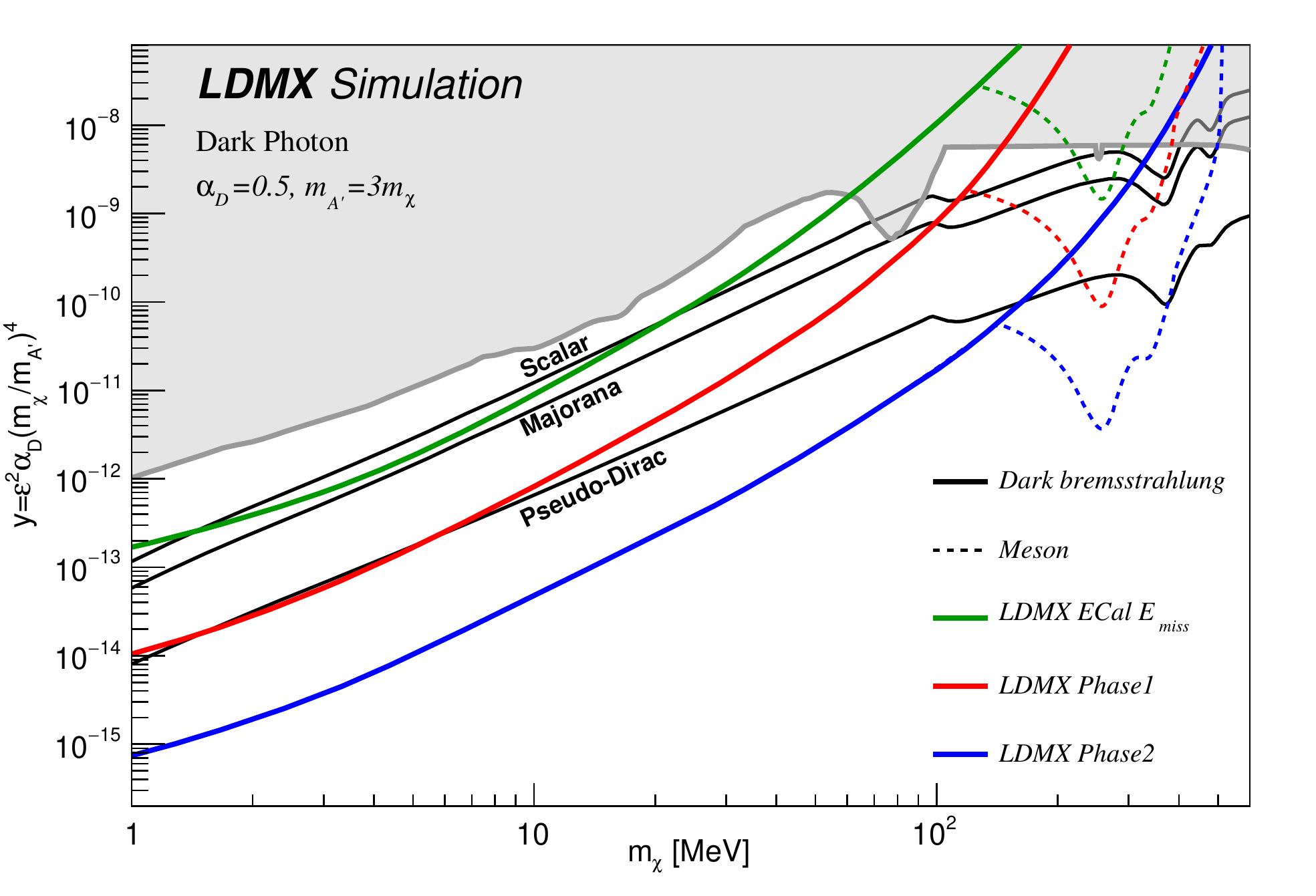}
    \caption{Projected sensitivity in the $y$ vs $m_{\chi}$ plane for a baseline LDMX run with $4\times10^{14}$~EoT with a 4~GeV beam energy (Phase 1) and $10^{16}$~EoT (Phase 2) with an 8~GeV beam energy. Projections for an early missing-energy analysis of $10^{13}$~EoT using the ECal as a target are also shown. 
    Backgrounds are assumed to be at the level of $<1$ event, supported by detailed simulation studies for the ECal missing-energy and 4~GeV missing-momentum analyses.
    Benchmark thermal relic targets are shown as black lines. 
    The grey region represents constraints from previous experiments.
    }
    \label{fig:reach}
\end{figure}

The combination of these selection criteria results in less than one predicted background event in a $4\times 10^{14}$~EoT sample, based on full GEANT4 simulations of events with multi-GeV ($>2.8$~GeV) photons at LDMX in a 4~GeV electron beam~\cite{Akesson:2019iul}. This performance would allow LDMX to probe thermal targets up to the O(100) MeV mass range with a 4~GeV beam energy, as indicated by Fig.~\ref{fig:reach}.

\subsection{Data-driven background estimation}
The data sample collected by the main physics trigger or supplemental triggers can be used to validate the ability of the baseline analysis to  reject background events. This sample can also be used for estimates of the background rates for the few events that remain in our signal region, by extrapolating events in data sidebands or by constraining closely related reactions in data. 

For example, the rate of neutral hadronic showers that leak through the HCal, and traverse the HCal modules undetected, can be constrained by measuring data sidebands defined by the HCal veto. Here, one can extract a punch-through probability by measuring the HCal energy spectrum, the hit amplitudes or the cluster sizes in the HCal modules and extrapolating to the low end of these distributions. Other neutron measurements can also be performed in data such as the rate of neutral hadronic showers as a function of the depth of the calorimeter, which can be used to extract a rate of single neutron backgrounds.

\begin{figure}
\centering
\includegraphics[width=0.6\textwidth]{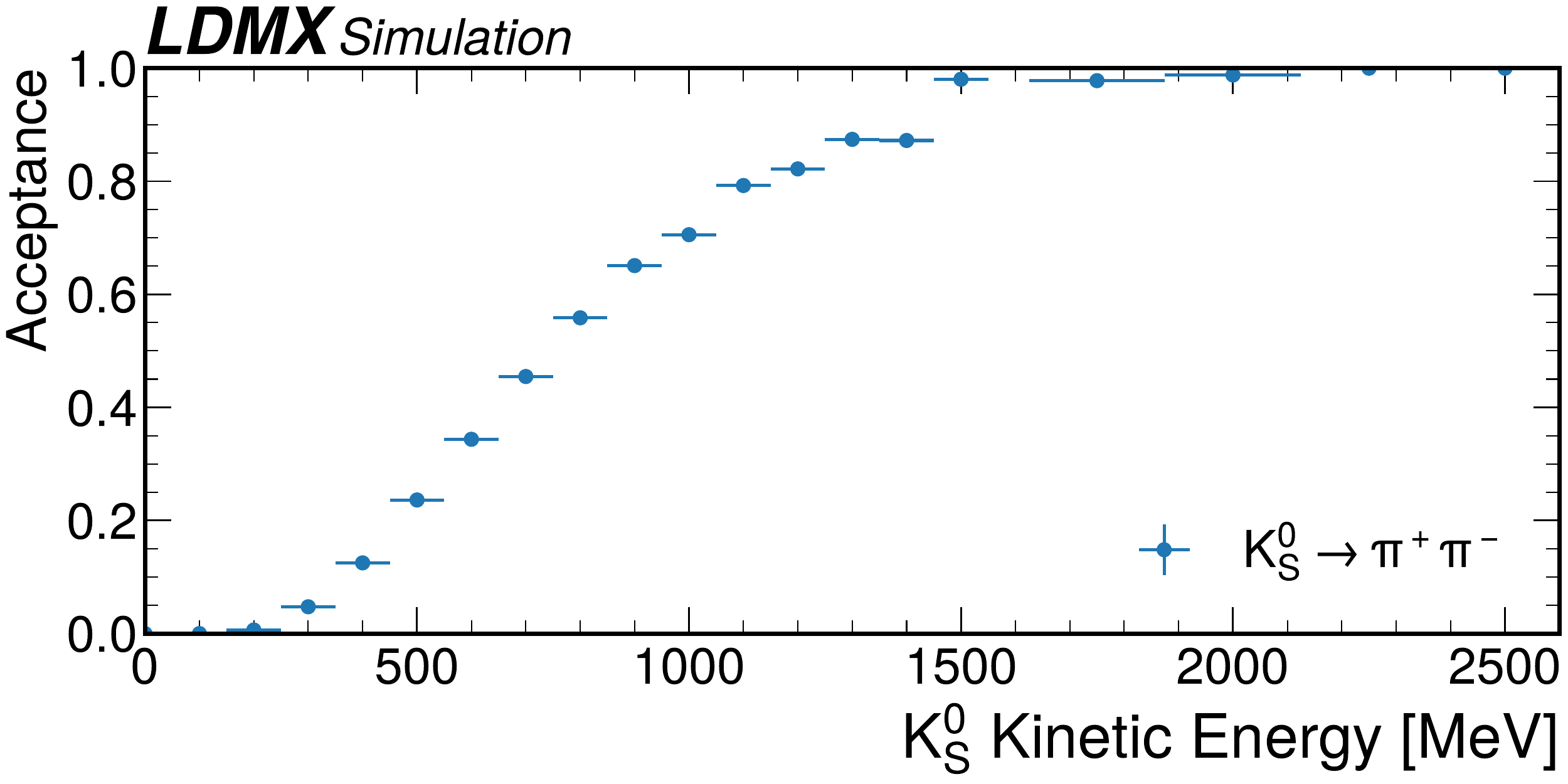}
\caption{The fraction of $K^0_S\to\pi^+\pi^-$ decays in which each daughter $\pi$ satisfies the detector acceptance, $\theta<40^\circ$, shown as a function of the $K^0_S$ kinetic energy. Here, $\pi$ candidates are reconstructed using particle-level four vectors and smearing their momenta by the estimated recoil tracker resolution.}
\label{fig:kaon}
\end{figure}

A different background process that can benefit from in-situ constraints is the production of charged kaons, most of which will leave no reconstructed track. A promising strategy to measure the rate of single $K$ production in photonuclear events exploits the visible decay channels of kaons that can be fully reconstructed in the detector. For example, $K^0_L$ events can be constrained by reconstructing $K^0_S$ decays into $\pi^+\pi^-$. These kaon candidates will have an approximate mass resolution of 30\%. Figure~\ref{fig:kaon} shows the acceptance of $K^0_S$ for photonuclear events produced in the target, where the majority of $q^2_{\gamma^*}$ is transferred to the kaon. For kaons with $E>1$~GeV its acceptance is larger than 50\%. Comparisons of data and simulated predictions of kaon production in the target will place constraints on the more experimentally-challenging process of kaon production within the ECal.

\subsection{Characterization of potential signals}

In the event that data is observed in excess of the small level of background expected to survive the vetoes described in Section~\ref{sec:fourGeV}, the question naturally arises of whether these events are consistent with a potential DM signal. One key discriminant is the mass of the (dark-)bremsstrahlung system, which, though invisible, is imprinted on the momentum of the visible recoil electron. Specifically, while background processes lead to generally soft and collinear photon emission that produces a low-\pt\ recoil $e^-$, massive DM mediators correspond to a large transverse kick and lower energy. The excellent energy and momentum resolution of LDMX thus allow for tight constraints on possible $A'$ masses.

The potential sensitivity of LDMX to the underlying signal model parameters using this method was investigated for the case of elastic scalar DM. Signal events were generated with a range of $A'$ masses from 0.4 to 2000~\,MeV, using the standard choices of $m_{A'}=3m_{\chi}$ and $\alpha_{D}=0.5$, which are used to derive 2-dimensional templates in recoil electron energy and \pt, as a function of $m_{A'}$. Thus, given a set of $n_{obs}$ events generated under a given signal hypothesis, the best-fit $A'$ mass can be extracted from a fit to the sample's recoil $E$ and \pt\ spectrum. At the same time, the observed number of signal events may be used to extract a range of allowed couplings $y$, which fully-determine the signal model's production cross section at LDMX.

\begin{figure}
\centering
\includegraphics[width=0.6\textwidth]{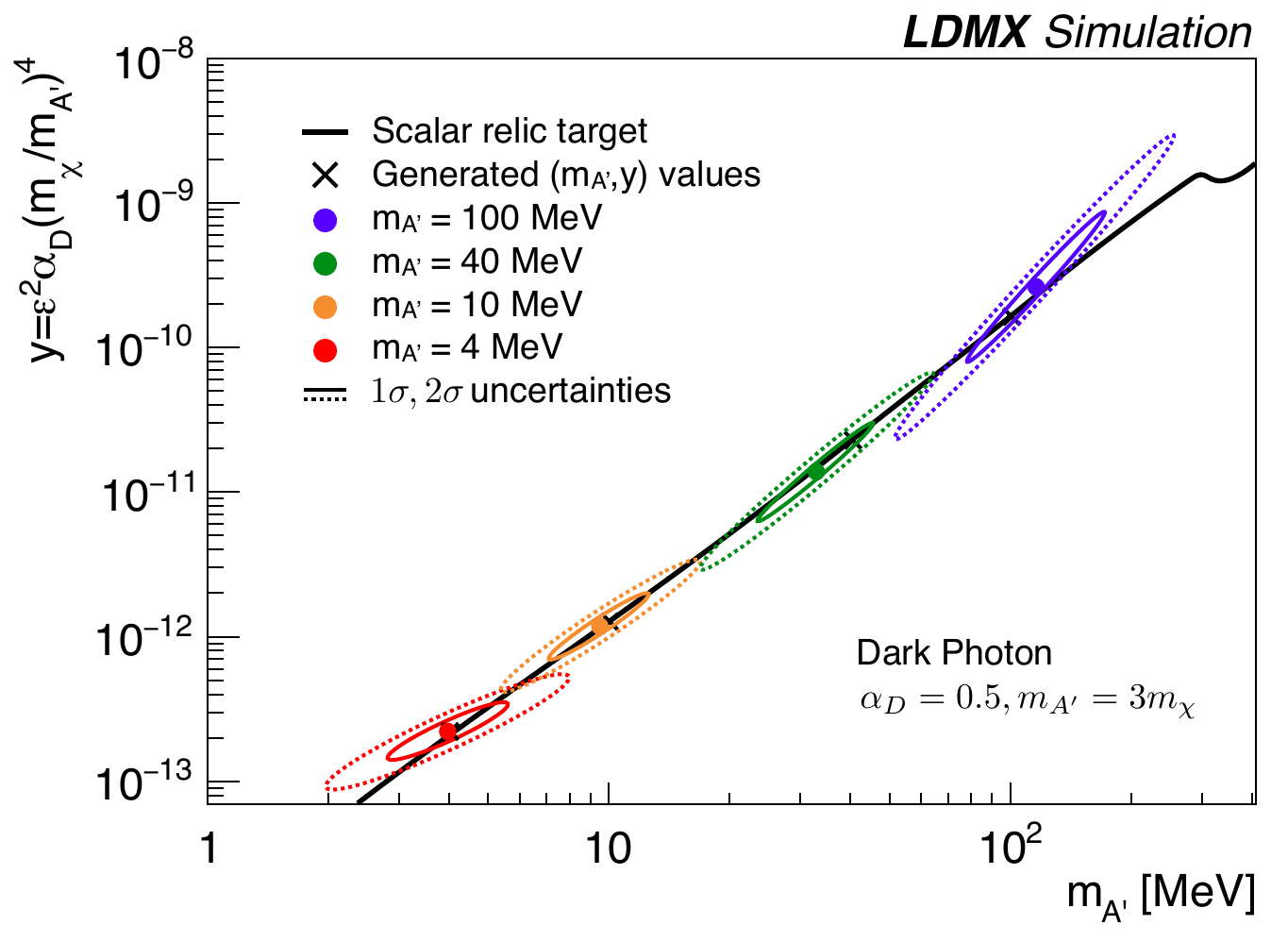}
\caption{Reconstructed mediator mass and couplings are shown for four test samples at $4\times10^{14}$~EoT.  The values of $m_{A'}$ and $y$ used to produce each sample are indicated by black crosses, while $1\sigma$ and $2\sigma$ error ellipses for the fit are denoted with solid and dashed lines, respectively.}
\label{fig:massReco}
\end{figure}

Figure~\ref{fig:massReco} shows the range of allowed signal parameters, considering scenarios where various $A'$ masses are realized in nature. Constraints are derived assuming $4 \times 10^{14}$~EoT and zero background, but are found to be robust even allowing for a small number of background events. In this scenario, the data would be sufficient to determine the dark photon mass to within approximately a factor of 2 at the 95\% confidence level, while $10^{16}$~EoT would improve this constraint to the sub-20\% level.

\subsection{ECal-as-Target analysis of early data}

The nominal DM search strategy at LDMX relies on the missing-momentum technique, based upon precise measurements of the electron momenta before and after interacting in the thin target. By contrast, a missing-energy search can be conducted to probe DM creation in the ECal, taking advantage of the large fraction of beam electrons that experience minimal interactions in the thin ($0.1X_0$) target. The methods are orthogonal and complementary, with the latter being of particular interest during early data-taking campaigns, due to the larger signal production cross section per~EoT.


A preliminary ECal-as-Target (EaT) analysis has been conducted, based on a DM signal generated through the developing shower of the beam electron as well as backgrounds enriched with EN and PN backgrounds, targeting an early data sample of $10^{13}$~EoT. The methodology is similar to that of the missing-momentum analysis described in Section~\ref{sec:fourGeV} in its use of ECal and HCal data to reject rare SM backgrounds. After the missing energy trigger, a BDT requirement on the ECal shower shape, an HCal activity veto, and dedicated ECal track rejection, less than one background event is expected for the $10^{13}$~EoT sample. This leads to a projected sensitivity that covers significant new territory across the $y$--$m_\chi$ plane, as shown in Figure~\ref{fig:reach}. Ultimately, results from this channel could be combined with the missing-momentum search, leading to further enhancements to the sensitivity of later phases of the experiment.

\subsection{Prospects with an 8~GeV beam}
\label{sec:eightGeV}

Searches for dark matter produced from 4\,GeV electron interactions have been the focus of previous sections, as this is the beam energy LESA plans to deliver in the initial stages of LDMX running. However, while 4\,GeV data will allow LDMX to access a significant new region of DM models, even higher energy beams ($\lesssim20$\,GeV) are projected to further extend the experiment's sensitivity, shown in Figure~\ref{fig:reach}. This is because signal yields increase at higher energies, while the rates for several challenging final states of photonuclear reactions scale as $1/E_\gamma^3$. Generally speaking, a higher beam energy makes it more difficult for SM reactions to fake the missing momentum signature. 
 
\begin{figure}
\centering
\includegraphics[width=0.98\textwidth]{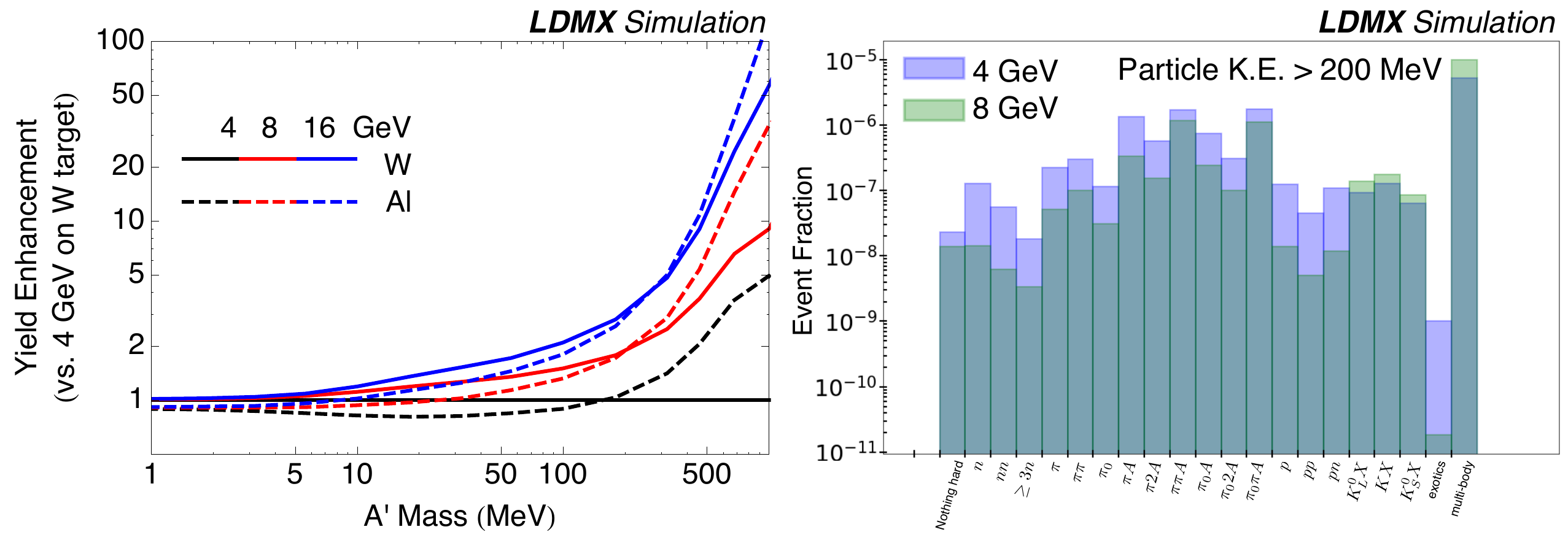}
\caption{
At left, DM production rates are shown as a function of beam energy, target composition, and mediator mass, relative to a 4~GeV beam with W target. At right, the composition of PN events in terms of the final state particle species is shown for 4 and 8\,GeV beams.
}
\label{fig:8gev}
\end{figure}

These effects are illustrated in Fig.~\ref{fig:8gev}: The yield enhancement as a function of dark photon mass is shown on the left. Different colours correspond to different beam energies, solid lines are for a tungsten target, dashed lines for an aluminum target. The baseline is a tungsten target and a 4\,GeV beam energy. The increase in yield is most pronounced for masses above 100\,MeV. The right hand side plot shows the fraction of photonuclear events that result in a specific final state, where particle multiplicities are counted considering particles with a kinetic energy above 200\,MeV. The blue histogram corresponds to 4\,GeV beam energy, the green one to 8\,GeV. It is found that almost all low-multiplicity final states -- which are the ones most difficult to catch -- occur less often by close to an order of magnitude, with the exception of final states containing kaons. 

Indeed, the bulk of the LDMX data will be collected at 8\,GeV after the upgrade of LCLS-II. In light of this, a simulation study of the rejection power for photon-induced backgrounds at 8\,GeV is being performed, following closely the methodology outlined for 4\,GeV above. Preliminary results indicate that the veto efficiency can be improved by an order of magnitude, while maintaining similar signal efficiencies as at 4\,GeV.  The resulting projected sensitivity to dark photons is shown in the Phase-2 projection of Figure~\ref{fig:reach}.

\subsection{Studies of multi-electron events}
\begin{figure}
\includegraphics[width=0.48\textwidth]{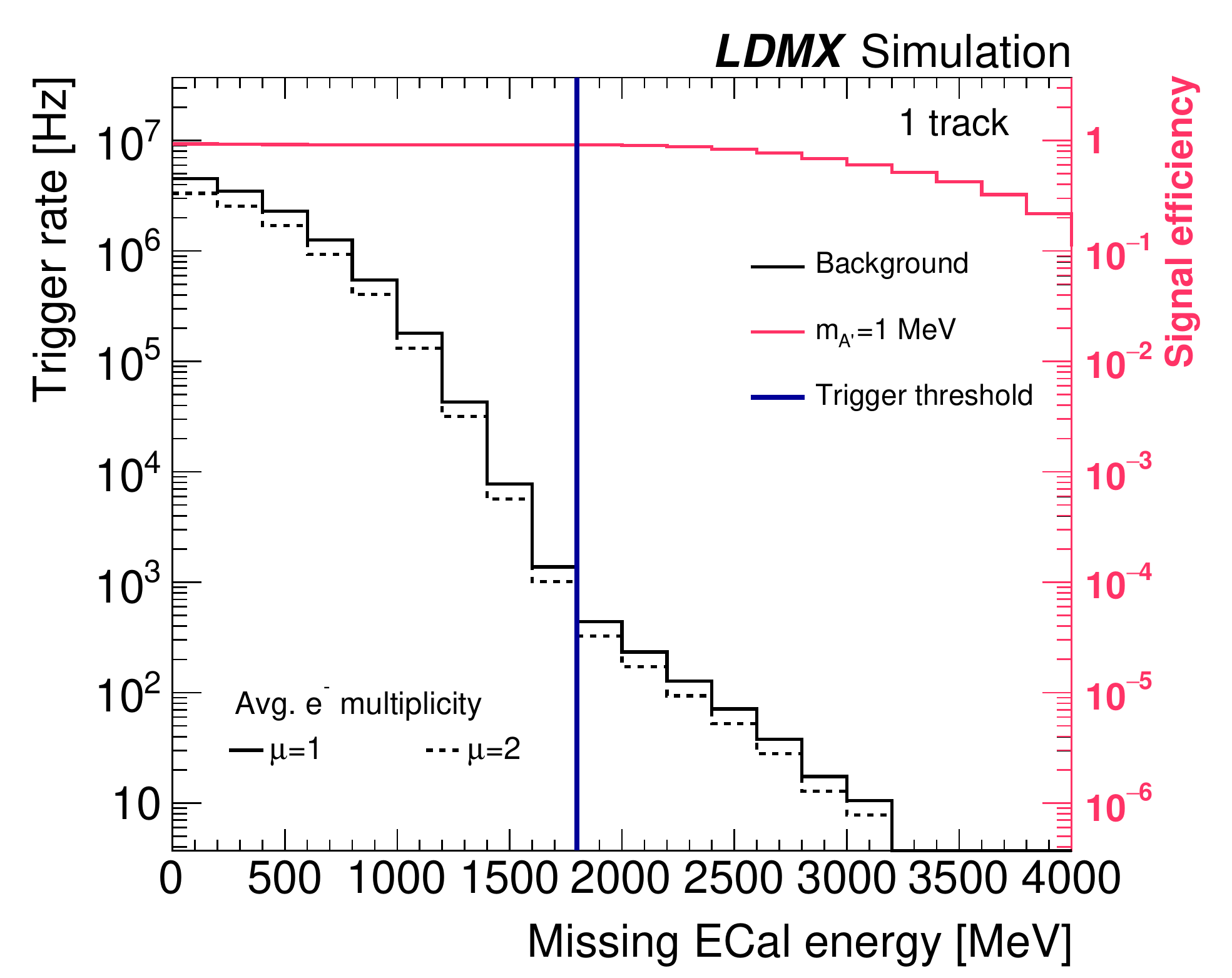}
\includegraphics[width=0.48\textwidth]{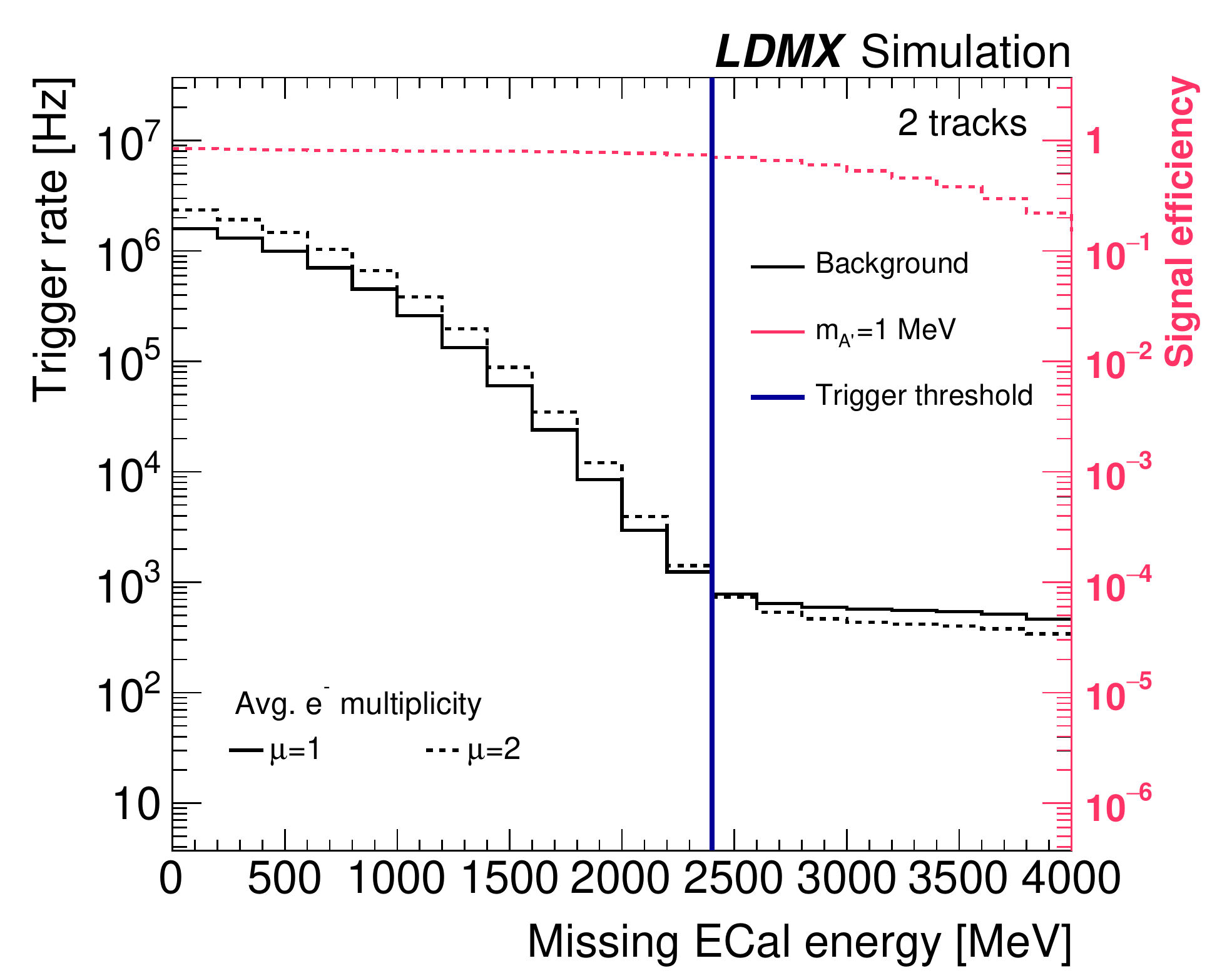}
\caption{
The trigger rate of accepted events is shown as a function of the total ECal energy, for events with one (left) and two (right) reconstructed tracks in the TS.
Each sample of simulated events contains a poisson-distributed number of simulated electrons in each time-sample with an average of either $\mu=1$ or 2. The total trigger rate is obtained by adding each $N_\text{track}$ category.
The efficiency for a dark photon signal with $m_{A'}=1$\,MeV is also shown for the same total ECal energy requirement.
}
\label{fig:multiEleTrigger}
\end{figure}

Efficient analysis of multiple incident beam electrons in a single time sample is a necessary capability to accumulate and exploit the Phase-2 sample of $10^{16}$~EoT. The presence of additional ``pileup'' electrons will lead to new challenges in triggering and reconstruction and, without suitable compensation, this may reduce the ability to separate the DM signal from a more challenging background. The missing-energy trigger will rely on a count of the number of incoming electrons ($n_e$) provided by the trigger scintillator system to determine the expected total energy of the incoming electrons. Accordingly, this missing-energy threshold may vary as a function of $n_e$ in order to optimize signal efficiency while keeping the overall trigger rate within the allocated bandwidth, as demonstrated in Figure~\ref{fig:multiEleTrigger}. Critically, an overestimate of $n_e$ would lead to an overestimate of incoming energy and thus produce a fake missing energy signal.

A triple-space-point tracking algorithm is used to count electrons with very low fake rate, exploiting positional and timing information from the trigger scintillator system. The fact that the three trigger scintillator pads are laid out to follow a beam electron's expected trajectory allows for fitting a constant to the hit positions (in a suitable coordinate system), which is a necessary constraint for making a track from only three points. Second, the time of flight from the first to the last of the three scintillator pads is greater than the time resolution of the system, allowing suppression of tracks formed with delayed hits. 


Beyond the task of designing a trigger strategy that maximizes the advantage of a high-current beam, dedicated reconstruction schemes will be required to analyze events with multiple interacting electrons. 
As the rate of multiple photonuclear interactions occurring in the same time sample is negligible, the task reduces to performing the analysis of Ref.~\cite{Akesson:2019iul} in the presence of a second more common (and easily rejected) electron interaction. While tracker and HCal vetoes require little-to-no modification, ECal reconstruction must be adapted to account for additional showers. The excellent segmentation of the ECal, together with the large size of the beamspot, enable an optimized multi-electron analysis to perform this task, the details of which are left to a future work.

\section{Searches for other light new physics}
\label{sec:otherNP}





Up to this point, projected sensitivities of missing-momentum searches have been cast in terms of dark photon models capable of explaining the observed DM abundance.
However, LDMX will have the capability to probe motivated and unconstrained regions of parameter space for a broader variety of models, including both visible and invisible signatures~\cite{Berlin:2018bsc}.
While the methods described in Section~\ref{sec:dm} will directly translate to many of these scenarios, dedicated visible searches will bring complementary coverage provided the associated technical challenges can be met.


\subsection{Further invisible signatures}

A series of relic targets were introduced in Section~\ref{sec:dm:intro}, given certain assumptions on the relative masses of the DM and mediator particles, and on the dark sector coupling.
While altering these assumptions can adjust DM production cross sections at LDMX for a given model of interest, generally this leads to minimal impact on the experimental methods.
When considering a wider class of mediators (such as scalars and axial-vectors), or even more exotic scenarios such as milli-charged DM, the missing-momentum signature at LDMX is unchanged, though the model details may leave an experimental imprint in the form of the recoil electron kinematics~\cite{Blinov:2020epi}.
In a similar vein, the invisible search will probe light $B-L$ gauge bosons, and scenarios where the DM is comprised of strongly-interacting massive particles (SIMPs).

Recently, it has been suggested that DM searches at NA64 and LDMX may be interpreted in terms of the invisible decays of the large number of light, unflavored mesons produced in the dump~\cite{Schuster:2021mlr}.
In particular, LDMX stands to improve over current limits on the branching ratios $\mathcal{BR}(V\to\chi\chi)$ by several orders of magnitude for $V=\rho,\omega,\phi$.
Unlike the signals considered above, in this case the DM is accessed through the decay of SM particles and not the production of a new mediator particle, which leads to a recoil electron \pt\ spectrum that is largely identical to the PN background.
This novel production channel can lead to tighter constraints for dark photon models than the bremsstrahlung mode at larger DM masses, as shown in Figure~\ref{fig:reach}.

\subsection{Visible Signatures}

In scenarios where the dark photon mass $m_{A'}$ is less than $2m_\chi$, the invisible decay channel exploited by the missing momentum search closes and visible decays of the $A'$ back to SM particles must be considered.

When $m_{A'} < m_\chi$, the relic abundance of this ``secluded dark matter'' is solely determined by dark sector interactions, dispensing with the convenient predictivity of the inverted mass hierarchy.
The off-shell and resonant regimes $m_\chi\lesssim m_{A'} \lesssim 2m_\chi$ require careful treatment: 
visible and invisible channels compete and DM annihilation can be enhanced in the early universe, allowing the viability of smaller values of $\epsilon$.
In these scenarios, weakly-coupled mediators present the possibility of searches for long-lived particles that travel macroscopic distances before decaying into SM particles.
Though optimized for a missing momentum search, LDMX is effectively a fully-instrumented, short baseline beam-dump experiment and can probe a variety of dark sector models in motivated and untouched parameter space through searches for displaced decays.

\begin{figure}
\centering
\includegraphics[width=0.6\textwidth]{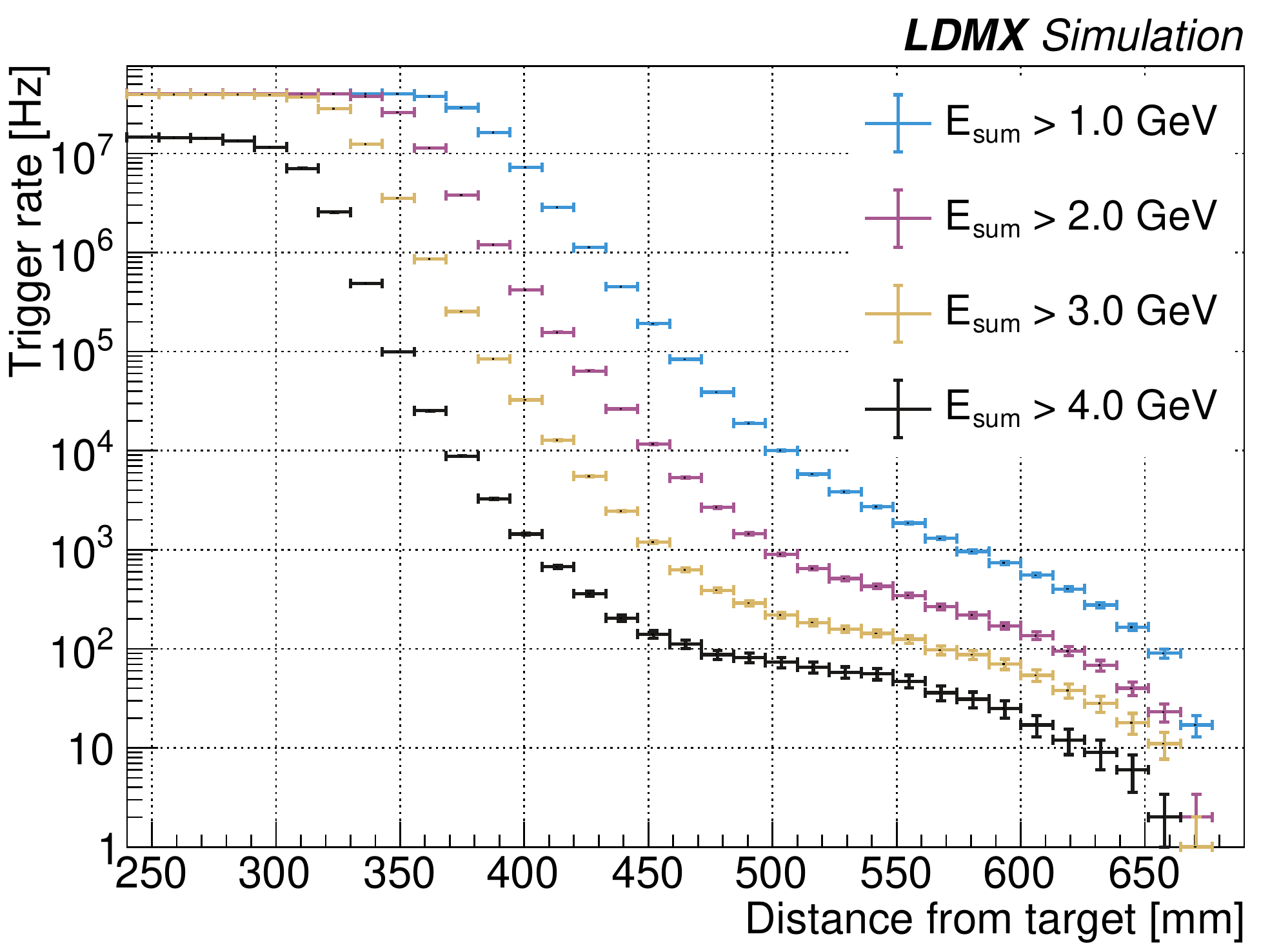}
\caption{
The rate of inclusive single-electron events passing a requirement on the total ECal energy reconstructed from layers beyond a particular distance $z$ from the target.
Requirements that select displaced deposits carrying a large fraction of the beam energy can be made to reduce rates to the order of 100~Hz.
}
\label{fig:visibleTrigger}
\end{figure}

It is possible to search for long-lived particles that decay into SM particles (including light quarks, charged leptons, and photons) by searching for displaced energy depositions in either the ECal or HCal. 
Because the total energy deposited in these decays can be similar in magnitude to the beam, the standard ECal missing-energy trigger is not optimal.
While decays in the HCal are effectively invisible to the missing-energy trigger (based on ECal information alone) the trigger strategy must be extended to improve sensitivity to short lifetimes that yield decays in the ECal. 
Figure~\ref{fig:visibleTrigger} illustrates the effect of requiring a large sum of energy to be deposited in the last $N$ layers of the ECal, where both the $E_\text{sum}$ requirement and first layer included in the sum are varied.
The approach eliminates most of the dominant backgrounds (beam electrons, bremsstrahlung conversion, and tridents) that deposit energy comparable to the beam at the front of the ECal while retaining good efficiency for displaced decays predicted for a range of unexplored models.

\begin{figure}
    \centering
    \includegraphics[width=0.45\textwidth]{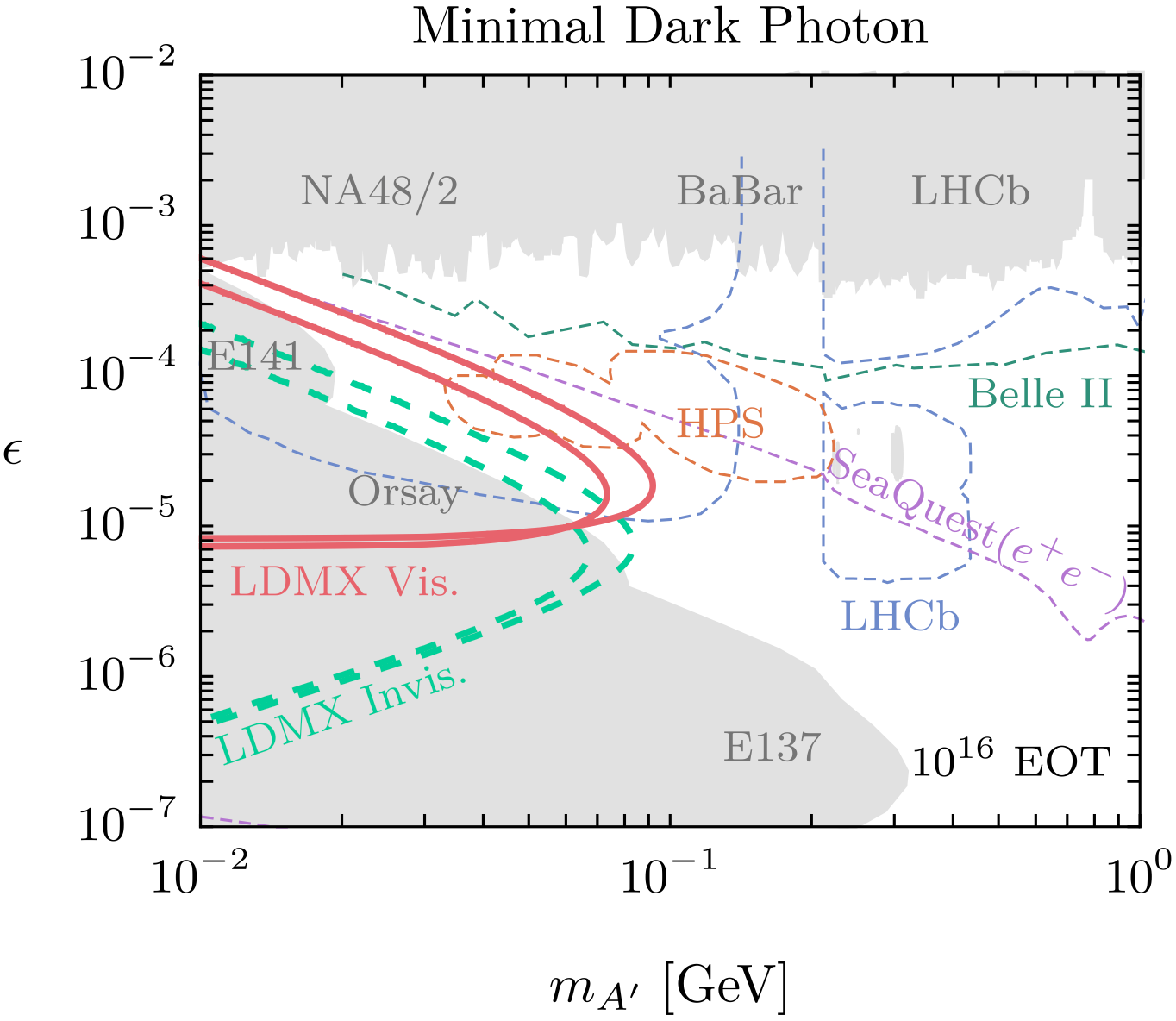}
    \includegraphics[width=0.45\textwidth]{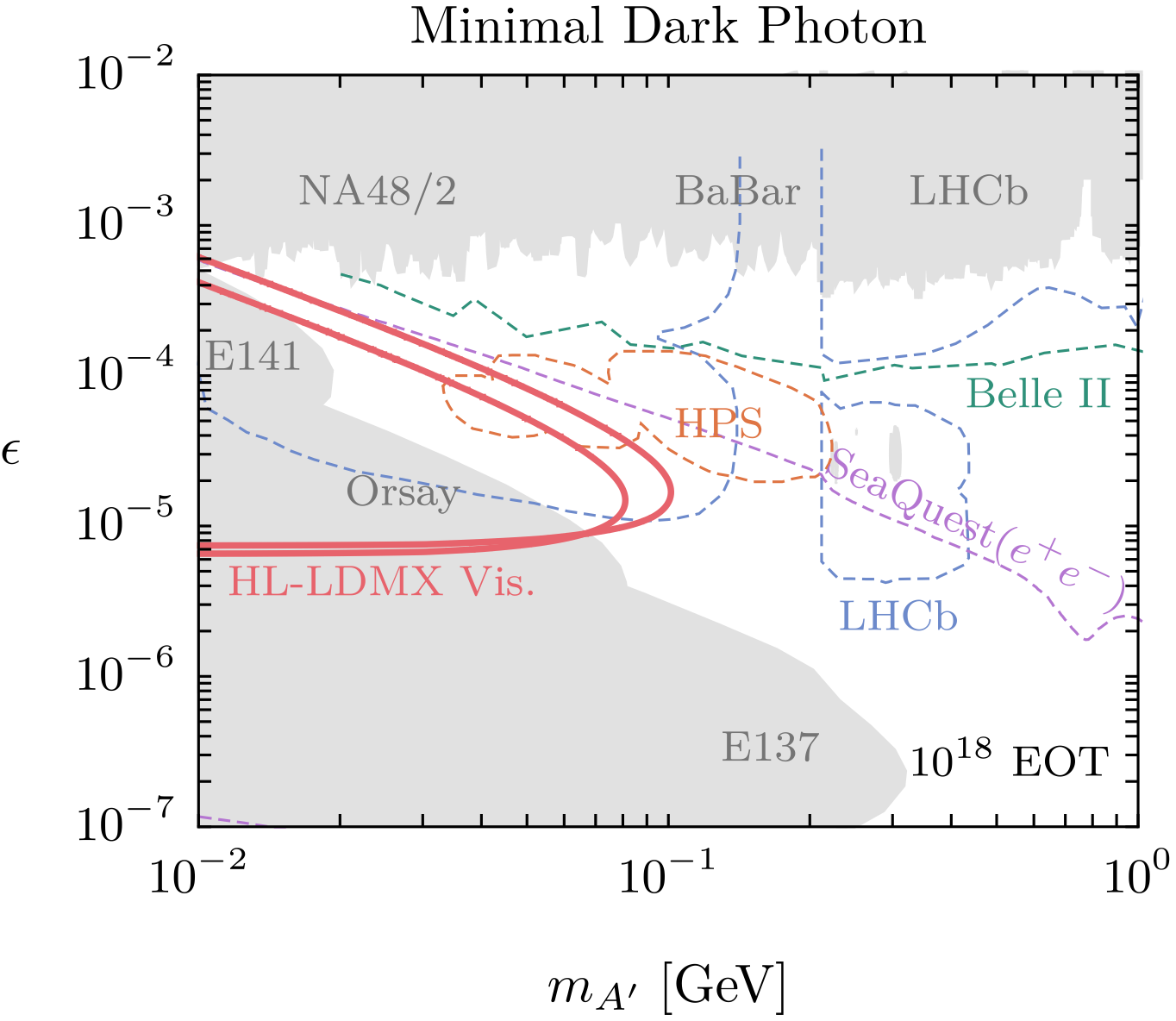}
    \caption{Estimates for the minimal dark photon model with visible decays at LDMX for an 8~GeV and 16~GeV beam energy. At left, projected sensitivities are shown for the missing-momentum and visible searches with $10^{16}$~EoT. At right, projections for a visible search are shown for a $10^{18}$~EoT high-luminosity run without tracker information. This figure is reproduced from Ref.~\cite{Berlin:2018bsc}}
    \label{fig:aprime_vis}
\end{figure}

Unlike the missing momentum search, a zero-background search for visible, displaced decays at LDMX is not generically feasible, due to the large number of events producing an EM shower near the beam energy and displaced SM interactions.
Many important backgrounds stem from the same physical processes as those in the missing momentum search, though in this case the concern is late energy deposition instead of undetected particles.
Exponentially falling late secondary photon conversion processes can be mitigated by considering signals farther downstream at the expense of reduced acceptance. 
More deeply penetrating processes, such as hard single-neutron final states and displaced decays of neutral kaons, can be mitigated using shower shapes in the ECal and transverse momentum of the recoil electron, though the specific strategy will be depend on the details of the model under study.
Ref.~\cite{Berlin:2018bsc} investigated the potential sensitivity of LDMX to unexplored minimal dark photon models through displaced searches, shown in Figure~\ref{fig:aprime_vis}.

As the visible signatures at LDMX do not explicitly rely on precise measurements of the recoil momenta, one can consider the possibility of dedicated tracker-less data-taking runs that might allow increased signal production rates through a combination of higher beam current and a thicker target.
The extended reach in a setup where the effective luminosity is increased by two orders of magnitude beyond the Phase-2 target is shown in Figure~\ref{fig:aprime_vis}(right).
However, even for visible signatures the transverse momentum of the recoil electron is useful to reject PN and late-conversion backgrounds, posing challenges of how to best leverage high-granularity ECal information to fully capitalize on high-luminosity runs.

\begin{figure}
    \centering
    \includegraphics[width=0.9\textwidth]{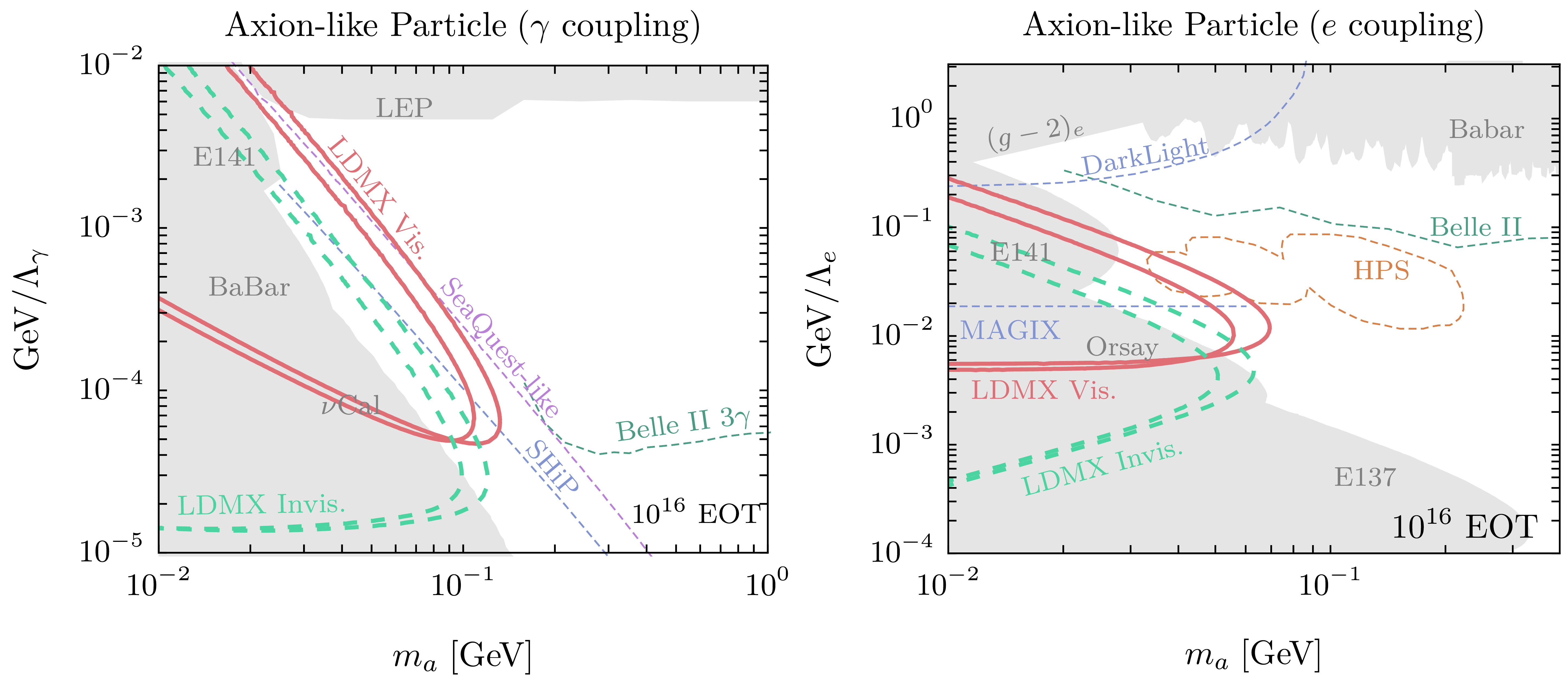}
\caption{The projected sensitivity of LDMX to the production of axion-like particles (ALPs) is shown as a function of the ALP mass and scale of the effective coupling to SM particles at a beam energy of 8~GeV and 16~GeV.
The left (right) panel shows the case where the ALP coupling to photons (electrons) is dominant. This figure is reproduced from Ref.~\cite{Berlin:2018bsc}}
    \label{fig:alp}
\end{figure}

Visible, displaced signatures are also predicted in models with axion-like particles (ALPs): light pseudo-scalar bosons coupled to the SM through dimension-5 effective interactions.
At beam dump energies, either the direct interaction with photon pairs or derivative coupling to electrons may be dominant depending on the details of the UV theory.
Thus in the electron-dominated case ALPs may be produced in analogy to the dark bremsstrahlung process, while in the photon-dominated case emission is via secondary photons.
Figure~\ref{fig:alp} shows projected limits on each scenario from LDMX using an 8 or 16~GeV beam with $10^{16}$~EoT~\cite{Berlin:2018bsc}.
For both couplings, the sensitivity of LDMX to previously-unconstrained models lies at the shortest lifetimes, as in the dark photon model.

A final possibility to consider is the prospect of semi-visible signatures, where stable DM is produced together with SM particles, potentially also with large displacements.
Such scenarios arise in models with richer dark sectors than the minimal dark photon case, such as inelastic DM and SIMP models.
As an example, production of new bound states of a dark QCD such as $\pi_D$ and $V_D$ in analogue to the SM mesons can lead to signatures such as $A'\to\pi_D (V_D\to\ell^+\ell^-)$ yielding displaced leptons and missing momentum.
A combination of analyses for visible and invisible signatures of New Physics at LDMX will constrain these models, though dedicated searches may bring enhanced sensitivity to this rich set of possible DM candidates.

\section{Measurements of electro-nuclear scattering}
\label{sec:en}

\begin{figure}[htp]
\centering
\includegraphics[width=0.43\linewidth]{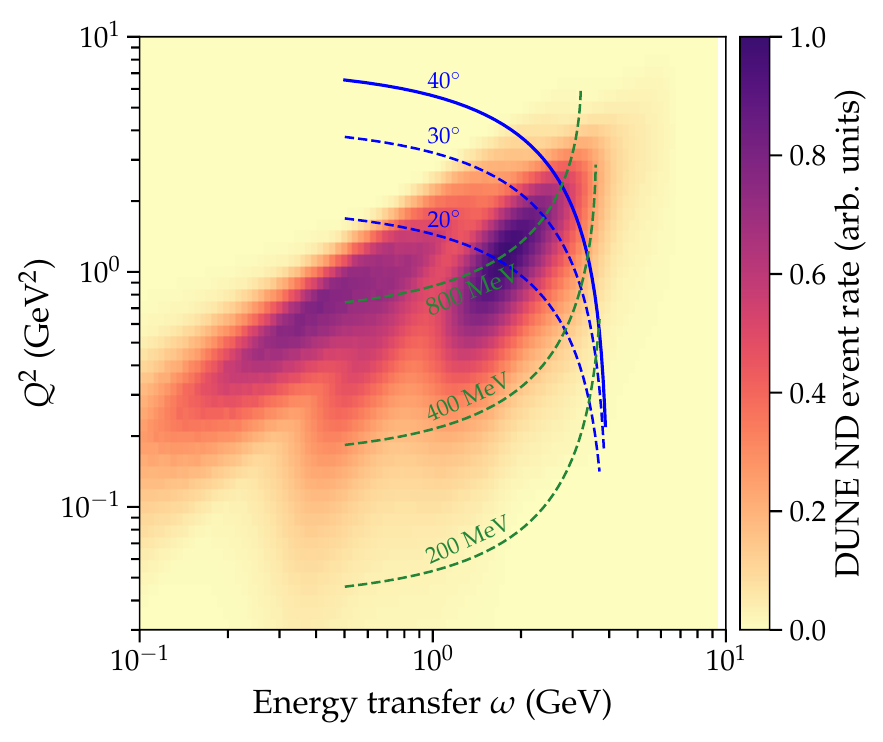}
\includegraphics[width=0.43\linewidth]{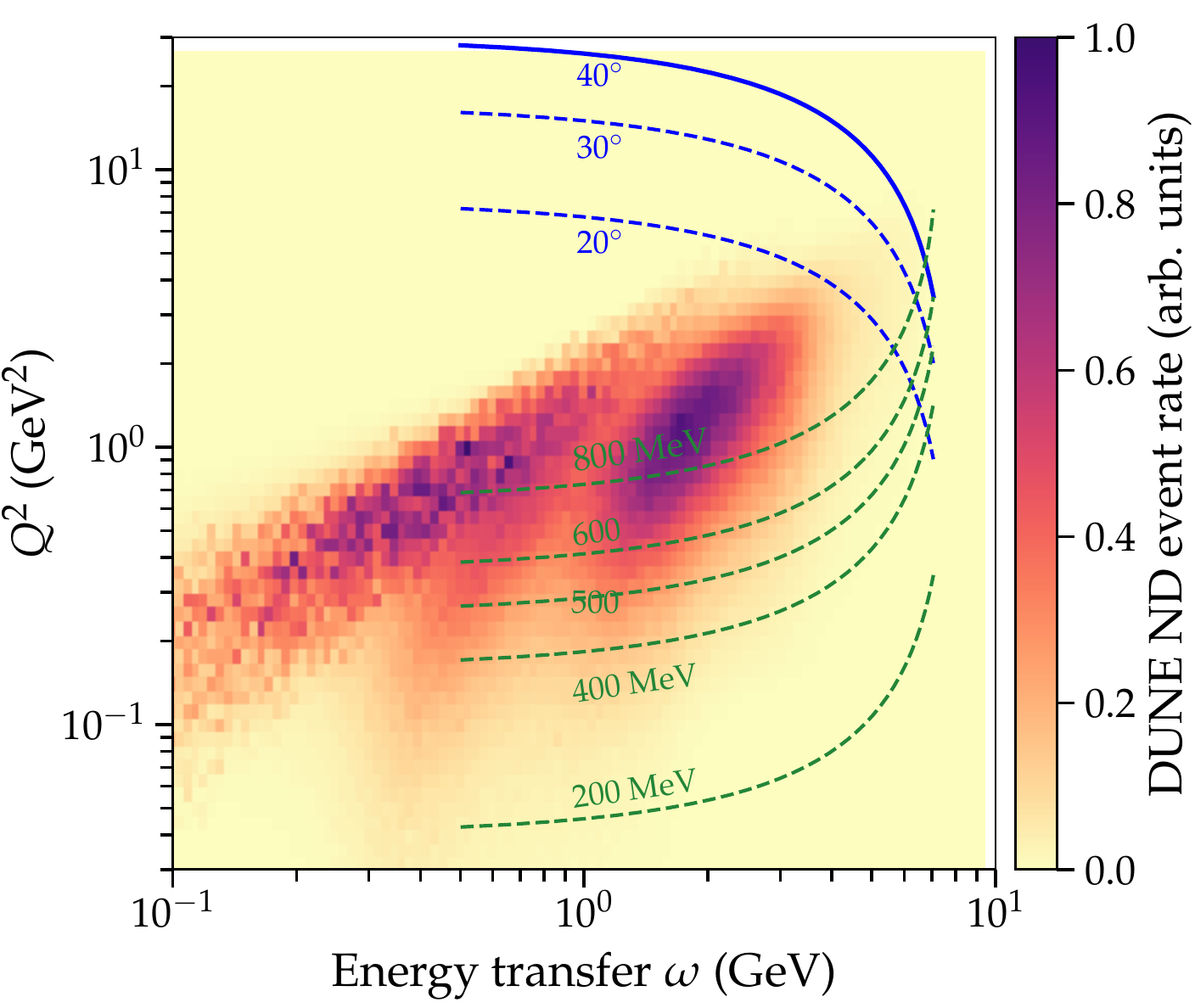}
\caption{Comparison of the phase space of neutrino-nucleon interaction events anticipated in the DUNE near detector in comparison with the LDMX detector acceptance. Both neutrino-nucleon and electron-nucleon interactions are specified as a function of the energy transfer from the incoming lepton to the hadronic system and $q^2$ of the interaction process. At left (right), the overlaid lines show LDMX acceptance assuming an 4\,GeV (8\,GeV) beam, corresponding to a set of maximum recoil electron polar angles and minimum transverse momenta.}
\label{fig:eN_phase_space}
\end{figure}

In addition to being a powerful tool to search for light dark matter, LDMX presents a unique opportunity to measure visible signatures of electron-nucleus (eN) interactions~\cite{Ankowski:2019mfd}.
Such interactions are of interest to the neutrino community, and the long-baseline experimental program in particular, because similar theoretical uncertainties impact both eN and neutrino-nucleus scattering.
The intrinsic challenge of modeling these effects is one of the largest uncertainties faced by current and future long baseline experiments~\cite{NOvA:2016vij,NOvA:2017ohq,NOvA:2018gge,Ankowski:2015kya}. 
Uncertainties related to neutrino interactions with nuclei stem from a number of sources, such as nucleon (Fermi) motion, binding energy, nucleon-nucleon correlations, and final-state interactions (whereby particles produced in the nucleus re-interact with the nucleus). 
A large fraction of these effects impact electron-nucleus interactions in the same way as neutrino-nucleus interactions.
In the case of neutrino oscillation experiments, these effects can lead to smearing and biases in the reconstructed neutrino energy, for which it is extremely difficult to correct. However, in LDMX, the incoming electron energy will be known to the percent level and the relationship between the electron energy and the observed final state can therefore be measured directly. The future DUNE experiment~\cite{DUNE:2020lwj,DUNE:2020ypp}  will utilize a wide-band neutrino beam with a peak energy of around 3~GeV. Figure~\ref{fig:eN_phase_space} compares the phase space for neutrino-nucleon scattering at DUNE to the expected acceptance of the LDMX experiment with 4~GeV and 8~GeV electron beams.


In general, the events of interest will have a final state consisting of a single electron and multiple hadrons, including protons, neutrons, pions, and other mesons. The LDMX design, including a recoil tracker, ECal and HCal, is expected to be able to observe and precisely reconstruct these multi-particle final states. Importantly, and unlike the majority of electron scattering experiments, the detector will in principle have acceptance down to scattering angles of zero.  It will also provide 2$\pi$ acceptance,  up to scattering angles of approximately 40$^\circ$.

\subsection{Recoil electron trigger}
\label{sec:en:trigger}
Due to the generally-visible detector signature of eN scattering, data collected with the main missing energy trigger must be supplemented with a new trigger strategy. A promising approach giving minimal bias on the physics process of interest is to identify events where the recoiling electron downstream of the thin target has significant momentum in the direction transverse to the beam (\pt). This requires a complete reconstruction of the electron shower in the electromagnetic calorimeter as well as the correlation of its position with the corresponding TS hits to infer the recoil momentum. Ensuring an acceptable trigger rate depends on the ability to reject similar signatures arising from Bremsstrahlung in the inclusive electron background, as well as detector resolution and multi-electron effects.


As tracking information is unavailable to the LDMX trigger, the electron \pt\ determination must rely on information from the ECal and TS alone. ECal trigger cells are comprised of groups of $3\times3$ sensors in a single hexagonal module, approximately 1\,mm$^2$ in area. The highest-energy trigger cells are transmitted to the global trigger system, where 2d clusters are formed in each ECal layer by adding nearest-neighbors to local maximum seed cells. Starting from the shower maximum, 2d clusters are propagated to the remaining layers to form 3d clusters, which are further refined to EM candidates by making loose cluster quality requirements. Finally the cluster can be combined with a TS track to form a trigger electron, whose \pt\ is obtained from the cluster energy and target-to-ECal displacement obtained from the cluster centroid and 2d hit coordinate at the target.

\begin{figure}
\centering
\includegraphics[width=0.8\textwidth]{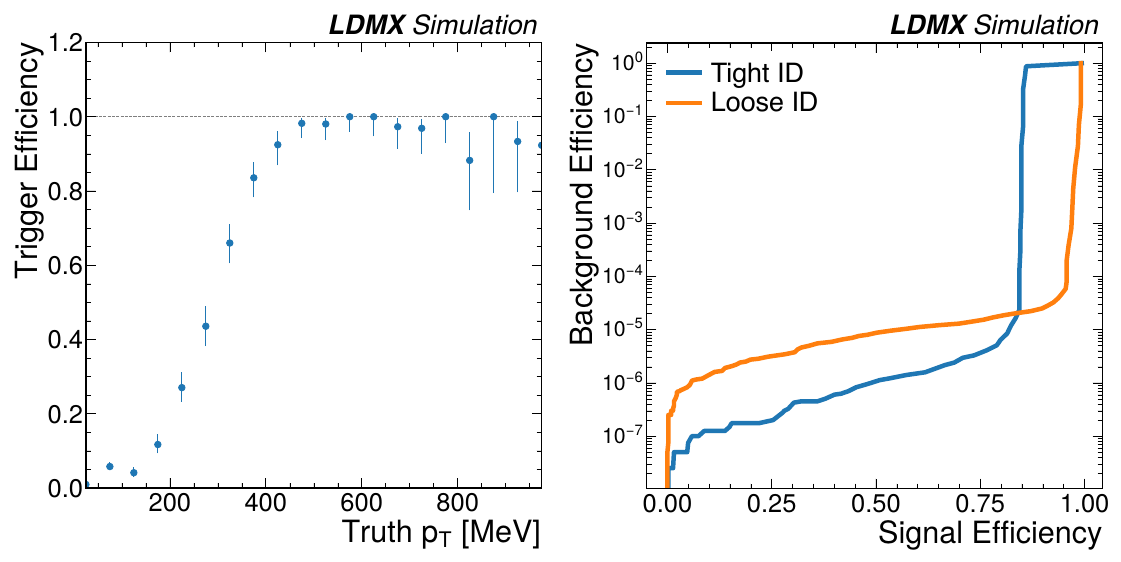}%
\caption{At left, the efficiency is shown for the high-$p_{T}$ electron trigger algorithm as a function of the truth electron $p_{T}$, obtained from a simulated samples of single electrons. At right, the signal efficiency measured in a sample of eN events with recoil  $\pt>400$ MeV is shown versus the background rejection in a sample of inclusive 4\,GeV single-electron events. Curves obtained by scanning the trigger electron \pt\ cut are shown for configurations with loose and tight requirements on the EM cluster quality.}
\label{fig:trigger}
\end{figure}

The efficiency for a trigger that selects events having one electron with $\pt>400$\,MeV is shown in Figure~\ref{fig:trigger}\,(left), plateauing at full efficiency. To explore the possibility of further lowering the trigger rate for a given \pt\ threshold, a tighter selection on the EM cluster quality is studied. Figure~\ref{fig:trigger}\,(right) shows that rates may be reduced up to an order in magnitude for the price of a modest reduction in efficiency to collect the eN signal. Multi-electron backgrounds are found to be negligible after a requirement on the total reconstructed ECal energy.

\subsection{Neutral hadron reconstruction}

Once recorded via the high-\pt\ electron tag at trigger level, a rich set of eN final states can be explored. These measurements can include inclusive cross sections, differential in the kinematics of the recoil electron, as well as probes of exclusive final states, which hold the potential to disentangle uncertainties on nuclear interactions originating from distinct effects.  In the case of the latter, accurate measurements of the multiplicity and energy spectra across all species of hadrons will enable tighter constraints to be placed on models of nuclear interactions. While the recoil tracker will allow for precise measurement of charged particles, and the ECal will capture electromagnetic decays, long-lived neutral hadrons pose a greater reconstruction challenge.

\begin{figure}
\centering
\includegraphics[width=0.48\textwidth]{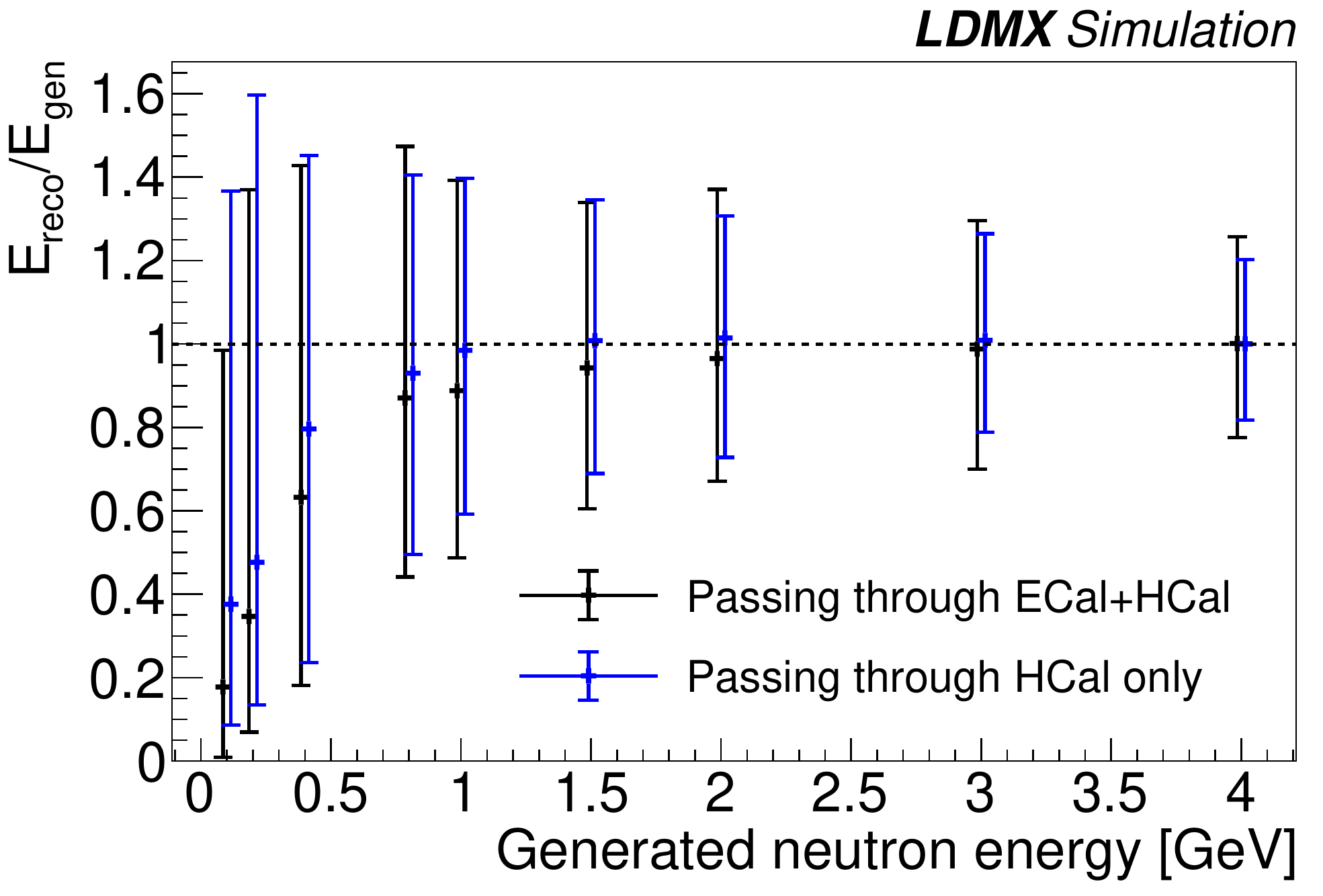}%
\includegraphics[width=0.48\textwidth]{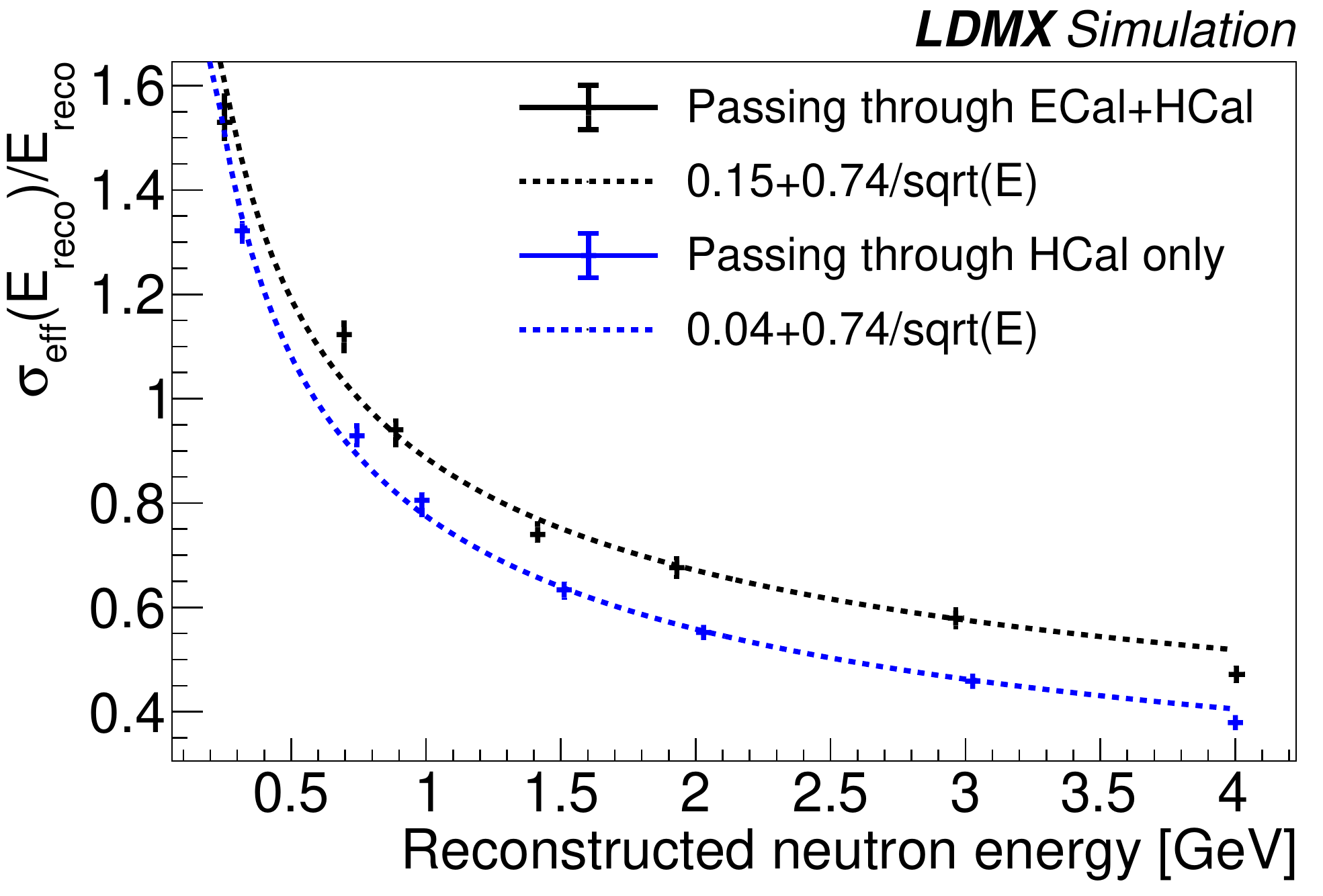}
\caption{At left, the response of the LDMX calorimeters to single neutrons is shown as a function of kinetic energy.  The median reconstructed energy is indicated by crosses for particles passing through either the complete ECal+HCal system or just the HCal alone.  Error bars indicate the 16 and 84\% quantiles of the reconstructed energy distribution.  At right, the effective RMS of the reconstructed energy distributions is shown, defined as the smallest interval containing 68\% of the distribution.}
\label{fig:neutron}
\end{figure}

The LDMX HCal is designed with the primary objective of serving as a hadronic veto for rare PN events where minimal energy is deposited in the ECal. In other words, the aim is to reconstruct or tag the presence of particles such as neutrons and $K_L^0$s with extremely high efficiency. It is interesting, however, to determine the capability to accurately determine the kinematics of the underlying neutral particles, for the purpose of constraining eN reactions. Accordingly, simulated samples of single particles have been used to determine the HCal response, including cases where particles are incident on the HCal only (as is the case for large-angle emissions), or the combined ECal+HCal system. Figure~\ref{fig:neutron} shows the single-particle response, as well as the corresponding energy resolution, where energies are summed across calorimeters. Optimizations taking into account the energy deposition across various calorimeter materials and longitudinal shower shapes have the potential to further improve these determinations, which we leave to future work.

\subsection{Particle ID with \texorpdfstring{$dE/dx$}{dE/dx}}
The silicon tracker may be used to infer the mass of charged particles that traverse the array of sensors, via their measured charge deposition. While choices related to the readout digitization are still being considered, preliminary studies of energy loss ($dE/dx$) have been carried out using simulated energy deposits. $dE/dx$ measurements are of significant interest to the electro-nuclear physics program, where $p/\pi$ discrimination at low energies will enable novel exclusive cross section measurements. These measurements can also allow to distinguish pions in data driven measurements (see Section~\ref{sec:dm}) or directly tag charged kaons in flight. 

\begin{figure}
\centering
\includegraphics[width=0.98\textwidth]{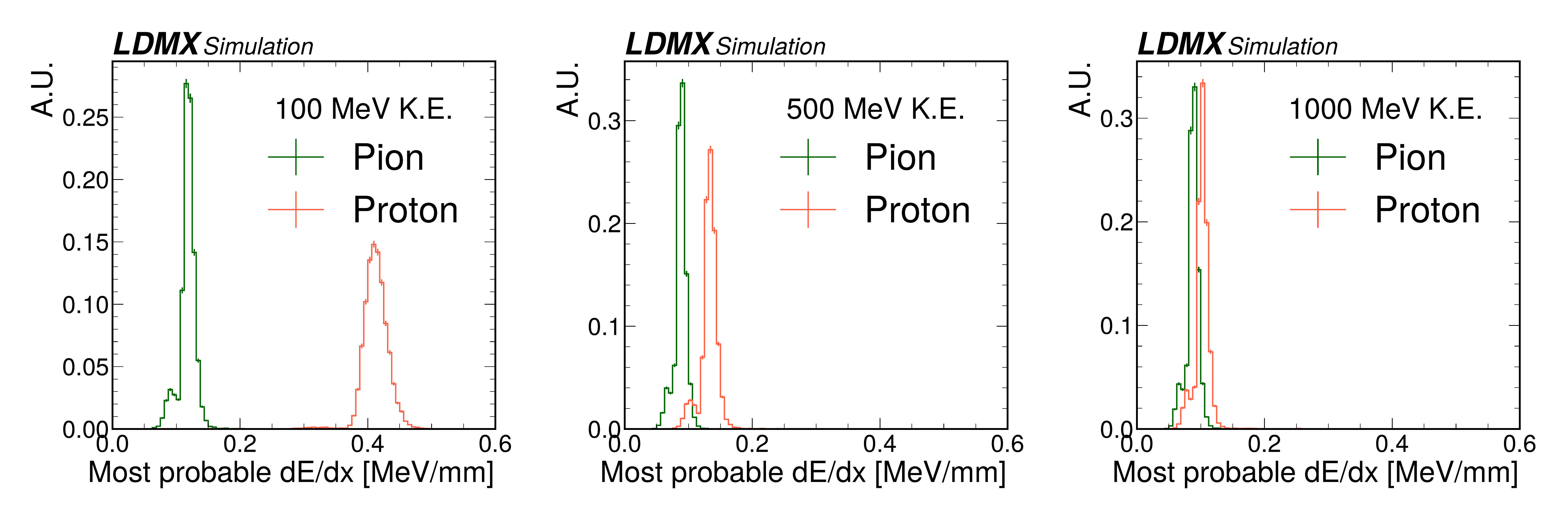}
\caption{Comparison of the most probable $dE/dx$ for protons and pions with kinetic energies of 100\,MeV (left), 500\,MeV (center), and 1\,GeV (right). Here, the most probable value of $dE/dx$ is inferred via template fits to the simulated energy deposits within the recoil tracker. }
\label{fig:pid}
\end{figure}

To obtain an event-level observable we fit the $dE/dx$ values corresponding to the simulated recoil tracker hits with a Landau distribution. From this fit we extract the most probable value of $dE/dx$ per event and account for contributions to the tail of the energy loss distribution by knock-on electrons or delta rays. Figure~\ref{fig:pid} compares the most probable values from the simulated energy deposition as a function of kinetic energy for charged pions and protons. Finally, we investigate the performance of other variables that can improve the $p/\pi$ discrimination, such as the average energy deposited or the sum of energy deposited divided by its distance where the maximum energy deposition happens. For kinetic energies $< 500$~GeV we expect approximately 30\% $\pi$ mis-tagging at 60\% $p$ identification efficiency. The mis-tag rate increases up to 50\% for kinetic energies near 1~GeV. We expect that a more realistic simulation of the recoil tracker improves the precision of these estimates.

\section{Prospects for future expansions of the LDMX physics program}
\label{sec:future}
While the LDMX apparatus as described in the previous sections will enable a robust program of New Physics searches and Standard Model measurements, one may consider the potential benefits of alternate run conditions, beam composition, and detector configurations. These present the potential to extend current capabilities along several avenues, the most exciting of which would be to enable precision measurements and characterization of a dark matter signal, should it be observed in the initial phases of running. Other possibilities include extending sensitivity to models of DM that are difficult to access with the Phase~2 dataset, expanding the phase space of electronuclear measurements, and probing invisible decays of light mesons.

\paragraph{Alternate beam configurations:}
In the event of the observation of a missing-momentum signal, it may be valuable to adjust the preparation of the beam to better characterize the mass scale, couplings, and spin structure of the new interaction. Background and signal production cross sections may experience variations of considerably different sizes due to adjustments in the beam energy, particularly for larger DM masses. While the currently planned LESA configuration does not support polarized beams, this development could prove a critical tool to probe the spin-parity of the mediator. Finally, variations of the missing-momentum search with muon beams could allow access to alternate models of light DM that provide solutions to the muon $g-2$ anomaly~\cite{Kahn:2018cqs}.

\paragraph{Varied target composition:}
The effects of exchanging thin W and Al targets on the dark matter production rate was previously shown in Section~\ref{sec:eightGeV}. While analogous high-$Z$ targets could be entertained, enhancements must be balanced with increased multiple scattering of the beam electrons. However, target variations may be especially useful in the event of an observed signal, due to different production scaling for the signal and background processes. In another vein, electro-nuclear measurements could benefit significantly from measuring interactions directly with the nuclei relevant for neutrino detectors.


\paragraph{Second-stage and track triggers:}
The LDMX trigger system as presently foreseen consists of a single stage of selection implemented in a system of custom electronics, tasked with providing a reduction in the data rate of approximately $10^4$. However, one may consider the benefits of adding a second stage in software, where tracker information may also be exploited. This could reduce the rejection factor required of the TS, allowing for higher DM signal efficiencies, and potentially enable the selection of novel visible signatures not possible with the current trigger design.

Similar advantages could be enabled if fast readout of the silicon strip tracking modules could be provided as input to a hardware track-finder for use in the trigger.
Such a system could improve electron counting and momenta determination to a level that may allow for runs with significantly increased luminosity. Additionally, this development would  for allow sizeable reductions in the threshold required for the high-\pt\ electron trigger used to collect eN interaction data.



\paragraph{Beyond $10^{16}$~EoT:}
The collection of a dataset beyond $10^{16}$~EoT would bring new challenges together with expanded physics potential. One intriguing target stems from the interpretation of the LDMX missing-momentum result as a limit on the invisible branching fraction of unflavored vector mesons~\cite{Schuster:2021mlr}. While Phase 2 would deliver world-leading limits, ultimate observation of the Standard Model $V\to\nu\bar\nu$ process would require an additional two-to-three orders of magnitude. Other methods of increasing the effective EoT through combinations of increased beam current and a thicker target may also be interesting for displaced searches, which do not explicitly rely on silicon tracking.







\section{Conclusions}
\label{sec:conc}
In this manuscript, we have reviewed the attractiveness of a light thermal relic as an explanation of the observed dark matter.
In the sub-GeV mass range, particle accelerators provide a generic probe of this possibility in fixed-target measurements.
We have introduced LDMX as a unique experimental apparatus to conduct a missing-momentum search for thermal relics, improving over current bounds on DM mediator interactions by orders of magnitude across a wide range of masses.
Detailed strategies have been presented to reduce SM backgrounds to negligible levels using state-of-the-art simulation tools, and in-situ methods have been proposed to constrain processes that may be a challenge to model precisely.
A path to scale these analyses to higher beam energies and currents was indicated.

In parallel, work establishing early detector prototypes has enabled numerous advances including, measurements of sensor responses, commissioning of the readout system, and validation of various design choices.
The completion of this work will naturally thrust LDMX into the impending construction phase equipped with a robust experimental design.

At the same time, efforts have expanded to study the broader physics program accessible at LDMX, including precision measurements of electro-nuclear interactions, rare meson decays, as well as searches for DM beyond the nominal missing-momentum search.
In the spirit of Snowmass, we have concluded by suggesting potential long-term avenues to enhance the physics potential of a future LDMX-like experiment.


\clearpage

\bibliography{references}

\begin{thebibliography}{10}

\bibitem{BRNReport}
{\em {Summary of the High Energy Physics Workshop on Basic Research Needs for
  Dark Matter Small Projects New Initiatives}}, 2019.

\bibitem{Battaglieri:2017aum}
Marco Battaglieri et~al.
\newblock {US Cosmic Visions: New Ideas in Dark Matter 2017: Community Report}.
\newblock In {\em {U.S. Cosmic Visions: New Ideas in Dark Matter}}, 7 2017.

\bibitem{Bjorken:2009mm}
James~D. Bjorken, Rouven Essig, Philip Schuster, and Natalia Toro.
\newblock {New Fixed-Target Experiments to Search for Dark Gauge Forces}.
\newblock {\em Phys. Rev. D}, 80:075018, 2009.

\bibitem{Izaguirre:2015yja}
Eder Izaguirre, Gordan Krnjaic, Philip Schuster, and Natalia Toro.
\newblock {Analyzing the Discovery Potential for Light Dark Matter}.
\newblock {\em Phys. Rev. Lett.}, 115(25):251301, 2015.

\bibitem{Izaguirre:2014bca}
Eder Izaguirre, Gordan Krnjaic, Philip Schuster, and Natalia Toro.
\newblock {Testing GeV-Scale Dark Matter with Fixed-Target Missing Momentum
  Experiments}.
\newblock {\em Phys. Rev. D}, 91(9):094026, 2015.

\bibitem{Alexander:2016aln}
Jim Alexander et~al.
\newblock {Dark Sectors 2016 Workshop: Community Report}.
\newblock 8 2016.

\bibitem{Akesson:2018vlm}
Torsten Åkesson et~al.
\newblock {Light Dark Matter eXperiment (LDMX)}.
\newblock 8 2018.

\bibitem{Raubenheimer:2018mwt}
Tor Raubenheimer, Anthony Beukers, Alan Fry, Carsten Hast, Thomas Markiewicz,
  Yuri Nosochkov, Nan Phinney, Philip Schuster, and Natalia Toro.
\newblock {DASEL: Dark Sector Experiments at LCLS-II}.
\newblock 1 2018.

\bibitem{Berlin:2018bsc}
Asher Berlin, Nikita Blinov, Gordan Krnjaic, Philip Schuster, and Natalia Toro.
\newblock {Dark Matter, Millicharges, Axion and Scalar Particles, Gauge Bosons,
  and Other New Physics with LDMX}.
\newblock {\em Phys. Rev. D}, 99(7):075001, 2019.

\bibitem{DUNE:2020lwj}
Babak Abi et~al.
\newblock {Deep Underground Neutrino Experiment (DUNE), Far Detector Technical
  Design Report, Volume I Introduction to DUNE}.
\newblock {\em JINST}, 15(08):T08008, 2020.

\bibitem{DUNE:2020ypp}
Babak Abi et~al.
\newblock {Deep Underground Neutrino Experiment (DUNE), Far Detector Technical
  Design Report, Volume II: DUNE Physics}.
\newblock 2 2020.

\bibitem{Hyper-Kamiokande:2018ofw}
K.~Abe et~al.
\newblock {Hyper-Kamiokande Design Report}.
\newblock 5 2018.

\bibitem{Hyper-Kamiokande:2022smq}
J.~Bian et~al.
\newblock {Hyper-Kamiokande Experiment: A Snowmass White Paper}.
\newblock In {\em {2022 Snowmass Summer Study}}, 3 2022.

\bibitem{Ankowski:2019mfd}
Artur~M. Ankowski, Alexander Friedland, Shirley~Weishi Li, Omar Moreno, Philip
  Schuster, Natalia Toro, and Nhan Tran.
\newblock {Lepton-Nucleus Cross Section Measurements for DUNE with the LDMX
  Detector}.
\newblock {\em Phys. Rev. D}, 101(5):053004, 2020.

\bibitem{Schuster:2021mlr}
Philip Schuster, Natalia Toro, and Kevin Zhou.
\newblock {Probing invisible vector meson decays with the NA64 and LDMX
  experiments}.
\newblock {\em Phys. Rev. D}, 105:035036, Feb 2022.

\bibitem{hgcalTDR}
{The Phase-2 Upgrade of the CMS Endcap Calorimeter}.
\newblock Technical report, CERN, Geneva, Nov 2017.

\bibitem{Bryngemark:2021}
Lene~Kristian Bryngemark, David Cameron, Valentina Dutta, Thomas Eichlersmith,
  Balazs Konya, Omar Moreno, Geoffrey Mullier, Florido Paganelli, Ruth
  Pöttgen, Fuzzy Rogers, and et~al.
\newblock Building a distributed computing system for ldmx.
\newblock {\em EPJ Web of Conferences}, 251:02038, 2021.

\bibitem{Marsicano:2018glj}
L.~Marsicano, M.~Battaglieri, M.~Bond\'\i{}, C.~D.~R. Carvajal, A.~Celentano,
  M.~De~Napoli, R.~De~Vita, E.~Nardi, M.~Raggi, and P.~Valente.
\newblock {Novel Way to Search for Light Dark Matter in Lepton Beam-Dump
  Experiments}.
\newblock {\em Phys. Rev. Lett.}, 121(4):041802, 2018.

\bibitem{Akesson:2019iul}
Torsten Åkesson et~al.
\newblock {A High Efficiency Photon Veto for the Light Dark Matter eXperiment}.
\newblock {\em JHEP}, 04:003, 2020.

\bibitem{Blinov:2020epi}
Nikita Blinov, Gordan Krnjaic, and Douglas Tuckler.
\newblock {Characterizing Dark Matter Signals with Missing Momentum
  Experiments}.
\newblock {\em Phys. Rev. D}, 103(3):035030, 2021.

\bibitem{NOvA:2016vij}
P.~Adamson et~al.
\newblock {First measurement of muon-neutrino disappearance in NOvA}.
\newblock {\em Phys. Rev. D}, 93(5):051104, 2016.

\bibitem{NOvA:2017ohq}
P.~Adamson et~al.
\newblock {Measurement of the neutrino mixing angle $\theta_{23}$ in NOvA}.
\newblock {\em Phys. Rev. Lett.}, 118(15):151802, 2017.

\bibitem{NOvA:2018gge}
M.~A. Acero et~al.
\newblock {New constraints on oscillation parameters from $\nu_e$ appearance
  and $\nu_\mu$ disappearance in the NOvA experiment}.
\newblock {\em Phys. Rev. D}, 98:032012, 2018.

\bibitem{Ankowski:2015kya}
Artur~M. Ankowski, Pilar Coloma, Patrick Huber, Camillo Mariani, and Erica
  Vagnoni.
\newblock {Missing energy and the measurement of the CP-violating phase in
  neutrino oscillations}.
\newblock {\em Phys. Rev. D}, 92(9):091301, 2015.

\bibitem{Kahn:2018cqs}
Yonatan Kahn, Gordan Krnjaic, Nhan Tran, and Andrew Whitbeck.
\newblock {M$^{3}$: a new muon missing momentum experiment to probe (g
  \ensuremath{-} 2)$_{\mu}$ and dark matter at Fermilab}.
\newblock {\em JHEP}, 09:153, 2018.

\end{thebibliography}

\end{document}